\documentclass[prd, twocolumn, showpacs, amsmath, preprintnumbers]{revtex4-1}
\usepackage[colorlinks=true,
            linkcolor=black,citecolor=black,urlcolor=black,
            breaklinks=true,
            pdftitle={A Student-t based filter for robust signal detection},
            pdfsubject={http://dx.doi.org/10.1103/PhysRevD.84.122004},
            pdfauthor={Christian Roever},
            pdfkeywords={robust signal detection, maximum likelihood detection, matched filter},
            bookmarks=false]{hyperref}
\usepackage{algcompatible} 
\usepackage{graphicx}      
\usepackage{dsfont}        
\providecommand{\realline}{\mathds{R}}
\providecommand{\complexnumb}{\mathds{C}}
\providecommand{\prob}{\mathrm{P}}
\providecommand{\imag}{\mathrm{i}}
\providecommand{\differential}{\mathrm{d}}
\providecommand{\expect}{\mathrm{E}}

\providecommand{\var}{\mathrm{Var}}

\providecommand{\invchisq}{\mbox{Inv-}\chi^2}
\providecommand{\re}{\mathrm{Re}}
\providecommand{\im}{\mathrm{Im}}
\DeclareMathOperator*{\argmax}{arg\,max}
\providecommand{\algAssign}{\;=\;} 
\renewcommand{\algorithmiccomment}[1]{// \textit{#1}}

\begin{document}

\title{A Student-t based filter for robust signal detection}
  \preprint{\texttt{LIGO-P1100103}}
  \preprint{\texttt{AEI-2011-060}}
\author{Christian R\"over}
  \email{christian.roever@aei.mpg.de}
  \affiliation{Max-Planck-Institut f\"{u}r Gravitationsphysik (Albert-Einstein-Institut) and Leibniz Universit\"{a}t Hannover, 30167~Hannover, Germany}

\begin{abstract}
  The search for gravitational-wave signals in detector data is often
  hampered by the fact that many data analysis methods are based on
  the theory of stationary Gaussian noise, while actual measurement
  data frequently exhibit clear departures from these assumptions.
  Deriving methods from models more closely reflecting the data's
  properties promises to yield more sensitive procedures.
  The commonly used \textsl{matched filter} is such a detection method
  that may be derived via a Gaussian model.
  In this paper we propose a generalized matched-filtering technique
  based on a \textsl{Student\mbox{-}$t$ distribution} that is able to
  account for heavier-tailed noise and is robust against outliers in
  the data. On the technical side, it generalizes the matched filter's
  least-squares method to an iterative, or adaptive, variation.
  In a simplified Monte Carlo study we show that when applied to
  simulated signals buried in actual interferometer noise it leads to
  a higher detection rate than the usual (``Gaussian'') matched filter.
\end{abstract}

\pacs{02.50.-r, 04.80.Nn, 05.45.Tp, 95.75.Wx}


\maketitle

\section{Introduction}
Since the existence of gravitational radiation was established as a
consequence from general relativity theory, a great amount of effort
has gone into the development of instruments and methods to detect
gravitational waves directly \cite{Thorne1987,Schutz1999}.
Gravitational waves (GWs) are notoriously weak compared to the sources
of noise in today's ground-based gravitational-wave detectors, and so
it takes both extraordinarily sensitive instruments as well as
sophisticated data analysis techniques to measure them.  The output of
an interferometric GW detector is essentially a time series of
nonwhite noise, and --- potentially --- a superimposed signal whose
exact waveform is determined by several parameters. Data analysis
aiming for GW detection hence requires filtering of time-series data
for rare, weak signals that are often of a known, parametrized shape.
Many commonly used approaches are based on \textsl{matched filtering}
the data.  The matched filter may be derived as a maximum-likelihood
(ML) detection method in the framework of a Gaussian noise model, but
more generally will actually be ML procedure for a wider class of
models.  While the method works remarkably well and is able to
discriminate weak signals from the noise, it commonly runs into
problems due to non-Gaussian or nonstationary behavior of the actual
instrument noise.  For example, the matched filter often is sensitive
to outliers or loud transient noise events in the data, which,
although showing little similarity with the signal sought for, also do
not look like plain noise either.  A lot of effort needs to go into
identifying such false alarms.

We propose a more robust procedure that is based on a
Student\mbox{-}$t$ distribution for the noise, as introduced in
Ref.~\cite{RoeverMeyerChristensen2011}.  Several motivations may be
used for introducing the Student\mbox{-}$t$ model; most obviously it
exhibits ``heavier tails'' and non-spherical probability density
contours, allowing one to accommodate outliers in the noise.
Alternatively, the model may also be seen as incorporating imperfect
prior knowledge of the noise spectrum, either because it is only
estimated to limited accuracy, or because it is varying over time.
Models of this kind are commonly used for robust \textsl{parameter
estimation}, but, as we will show in the following, the model also
exhibits a better performance for \textsl{detection} purposes when the
assumption of stationary Gaussian noise is violated.  We expect the
proposed filtering method to be useful in other signal-processing
contexts as well.

In Sec.~\ref{sec:GaussFilter} we will first derive the usual matched
filter from a Gaussian noise model. In Sec.~\ref{sec:StudentTFilter}
we introduce the Student\mbox{-}$t$ model, elaborate on the motivation
for its use as well as point out the differences from the Gaussian
model, and derive the analogous filtering procedure.  In
Sec.~\ref{sec:example} we report on a case study using real detector
data and simulated signals to show that here the Student\mbox{-}$t$
based filter indeed yields a better detection rate. We close with some
concluding remarks.

\section{Gaussian matched-filtering}\label{sec:GaussFilter}
\subsection{General}
A \textsl{matched filter} may be derived in different ways, for
example based on considerations of the residual sum-of-squares (or
\textsl{power}) decomposition, without reference to a more specific
noise model \cite{Turin1960}; however, here we will concentrate on a
derivation via the assumption of stationary Gaussian noise and the
Whittle likelihood. This will allow us to easily generalize the usual
matched-filtering method to the case of Student\mbox{-}$t$ distributed
noise in the following.
%
%
It is important though to keep in mind that the matched filter is not
necessarily connected to the assumption of Gaussian noise.  When we
say ``Gaussian matched filter'', this is meant to refer to its
derivation and interpretation in the Gaussian context.

\subsection{The Gaussian noise model}\label{sec:GaussModel}
In order to implement the assumption of stationary, Gaussian noise
residuals, the \textsl{Whittle likelihood} approximation is commonly
utilized
\cite{ChoudhouriGhosalRoy2004a,Finn1992,RoeverMeyerChristensen2011}.
In the Whittle approximation, signal and noise time series are treated
in their Fourier-domain representation.  The explicit assumption being
made on the noise~$n(t)$ is that its discrete Fourier
transform~$\tilde{n}(f)$ is independently Gaussian distributed with
zero mean and variance proportional to the power spectral density
(PSD),
\begin{equation}\label{eqn:GaussianModel}
  \var\Bigl(\re\bigl(\tilde{n}(f_j)\bigr)\Bigr)
  = 
  \var\Bigl(\im\bigl(\tilde{n}(f_j)\bigr)\Bigr)
  = 
  \textstyle\frac{N}{4\Delta_t}S_1(f_j),
\end{equation}
where $f_j$ is the $j$th Fourier frequency, $S_1(f_j)$ is the
corresponding one-sided power spectral density, and $j=0,\ldots,N/2$
indexes the Fourier frequency bins. An explicit definition of the
Fourier transform conventions used here is given in the appendix.

For some measured data~$d(t)$ one then commonly assumes a parametrized
signal~$s_\theta(t)$ with parameter vector~$\theta$ and additive
Gaussian noise with a known 1-sided power spectral density~$S_1(f)$:
\begin{equation}
  d(t)=s_\theta(t)+n(t)
  \quad\Leftrightarrow\quad
  \tilde{d}(f)=\tilde{s}_\theta(f)+\tilde{n}(f)  
\end{equation}
(i.e., additivity holds in both time and Fourier domains).
The corresponding likelihood function then is given by
\begin{eqnarray}
  p\bigl(d|\theta\bigr)
  &\propto&
  \exp\biggl(- {\textstyle\frac{1}{2}}\sum_j \frac{|\tilde{d}(f_j)-\tilde{s}_\theta(f_j)|^2}{\frac{N}{4\Delta_t}\,S_1(f_j)} \biggr) \label{eqn:GaussianLikelihood}
\end{eqnarray}
\cite{RoeverMeyerChristensen2011}.

\subsection{Likelihood maximization}\label{sec:GaussLikeliMaxi}
\subsubsection{ML detection and the profile likelihood}
If there were no unknown signal parameters to the signal model (like
time-of-arrival, amplitude, phase,...), then, according to the
\textsl{Neyman-Pearson lemma} \cite{MGB}, the optimal detection
statistic would be the likelihood ratio between the ``signal'' and
``no-signal'' models.  Once there are unknowns in the signal model, a
common approach is to use a \textsl{generalized Neyman-Pearson test
statistic}, that is, the \textsl{maximized} likelihood ratio, where
maximization is carried out over the unknown parameters \cite{MGB}.
While this is in general not an optimal detection statistic, this
\textsl{ad~hoc} approach is often efficient and effective.  Such a
maximum likelihood (ML) detection approach is closely related to ML
estimation, as either way the parameter values maximizing the
likelihood will need to be derived.  In case of a Gaussian noise model
as in~(\ref{eqn:GaussianLikelihood}), maximization of the likelihood
is equivalent to minimizing a weighted sum-of-squares, i.e., a
weighted least-squares approach.

It should be noted that in a Bayesian reasoning framework, the
detection problem would be approached via the \textsl{marginal
likelihood} rather than the maximized likelihood
\cite{BDA,Berger}. The marginal likelihood is the expectation of the
likelihood function with respect to the prior distribution, and both
marginal and maximized likelihood may be equivalent for a certain
choice of the prior distribution.  One can show the marginal
likelihood to be optimal for any particular choice of prior
distribution, while the maximized likelihood in general is not (see
e.g.\ \cite{PrixKrishnan2009,Searle2008}). Maximization of the
likelihood on the other hand is commonly much easier computationally.

As will be seen in the following, it is often convenient to divide the
parameter vector into subsets, as it may be possible to analytically
maximize the likelihood for fixed values of some parameters over the
remaining parameters. This maximized \textsl{conditional} likelihood
as a function of a subset of parameters is also called the
\textsl{profile likelihood}. If a profile likelihood is given,
likelihood maximization may be reduced to maximizing over the
remaining lower-dimensional parameter subspace.  As an example,
consider a signal having three free parameters: amplitude, phase, and
time of arrival. If likelihood maximization can be done analytically
over amplitude and phase for any given arrival time, this results in a
profile likelihood that is a function of time. The likelihood's
overall maximum then may be computed via a numerical brute-force
search of the profile likelihood over the time parameter.

\subsubsection{Why care about linear models?}\label{sec:WhyLiMo}
In signal processing in general, and in GW data analysis in
particular, the signals of interest are commonly parametrized (among
other additional parameters) in terms of an amplitude and a phase
parameter. Consider for example a simple sinusoidal signal of the form
\begin{eqnarray}\label{eqn:sinusoidExample}
  s_{A,\phi,f}(t) &=& A\,\sin(2\pi f t + \phi)\\
       &=& \beta_{\mathrm{s}} \,\sin(2\pi f t) + \beta_{\mathrm{c}} \,\cos(2\pi f t)
\end{eqnarray}
which instead of amplitude~$A$ and phase~$\phi$ may equivalently be
parametrized in terms of sine- and cosine-amplitudes
$\beta_{\mathrm{s}}$ and $\beta_{\mathrm{c}}$.  Other examples of
signal models given in terms of linear combinations are the singular
value decomposition approach used e.g.\ in
\cite{RoeverEtAl2009,CannonEtAl2010} or the transformation of antennae
pattern effects into four amplitude parameters in the derivation of
the $F$\mbox{-}statistic \cite{JaranowskiKrolakSchutz1998}.  A linear
model formulation will turn out convenient in the following, as a
linear (or conditionally linear) model will allow us to perform
(conditional) likelihood maximization analytically.

\subsubsection{The general linear model}
Consider a \textsl{linear} model for the data, i.e.,
\begin{equation}
  y = X \beta + \epsilon
\end{equation}
where $y$ is a $N$-dimensional data vector, $X$ is a $(N\times
k)$-matrix, $\beta$ is a $k$-dimensional parameter vector, and
$\epsilon$ is an $N$-dimensional vector of error terms.  The
errors~$\epsilon$ are assumed to be Gaussian distributed with mean
zero and some covariance matrix~$\Sigma$.

In the above signal-processing context, $y$ and $\epsilon$ are the
$N$\mbox{-}dimensional vectors of re-arranged real and imaginary parts
of Fourier-domain data~($\tilde{d}$) and noise~($\tilde{n}$), the
signal~$s_\theta$ is given by a linear combination of the columns of a
matrix~$X$ according to the parameter vector~$\beta$, and the noise
covariance~$\Sigma$ is a diagonal matrix defined
through~(\ref{eqn:GaussianModel}).

The Gaussian likelihood function is characterized by
\begin{eqnarray}
  p(y|\beta) 
  &\propto&
  - (y-X\beta)^\prime \,\Sigma^{-1}\, (y-X\beta).
\end{eqnarray}
In the linear model, the likelihood may be maximized analytically, and
the ML~estimator for the unknown parameter vector~$\beta$ is given by
\begin{equation}
  \hat{\beta} 
  \;=\;
  (X^\prime \Sigma^{-1}X)^{-1} X^\prime \Sigma^{-1} y
\end{equation}
\cite{BDA,NeterEtAl}.

In the models of concern here, estimation is simplified by the fact
that the noise covariance~$\Sigma$ is a diagonal matrix
(\ref{eqn:GaussianModel}) so that its inverse again is diagonal.  In
addition, here we add the common assumption that the vectors spanning
the signal manifold, the columns of~$X$, are orthogonal. A
non-orthogonal basis~$X$ would complicate the procedure slightly; see
e.g.~\cite{JaranowskiKrolakSchutz1998}.  Under these conditions, the
pivotal quantities for ML estimation and detection are
\begin{eqnarray}
  b_j 
  &=&
  X_{\cdot,j}^\prime\,\Sigma^{-1}\, y
  \;=\;
  \sum_{i=1}^{N}\frac{x_{i,j}\, y_i}{\sigma_i^2}
  \qquad\mbox{and}\label{eqn:bj}
  \\
  c_j 
  &=& 
  X_{\cdot,j}^\prime\,\Sigma^{-1} \,X_{\cdot,j}
  \;=\;
  \sum_{i=1}^{N}\frac{x_{i,j}^2}{\sigma_i^2}\label{eqn:cj},
\end{eqnarray}
i.e., the quadratic forms, or inner products, involving the $j$th
basis vector ($j$th column of~$X$) with the data vector~$y$, and with
itself.  The elements of the parameter vector's ML
estimate~$\hat{\beta}$ are then given by
\begin{eqnarray}
  \hat{\beta}_j
  &=& 
  \frac{b_j}{c_j},
\end{eqnarray}
the maximized likelihood ratio vs.\ the no-signal model is given by
\begin{eqnarray}\label{eqn:maxLrLimo}
  \log\biggl(\frac{p(y|\hat{\beta})}{p(y|\vec{0})}\biggr)
  &=& 
  \sum_{j=1}^k\frac{b_j^2}{2\,c_j},
\end{eqnarray}
and the fitted values are given by
\begin{eqnarray}
  \hat{y}
  &=& 
  X\hat{\beta}
  \;=\; 
  \sum_{j=1}^k\hat{\beta}_j X_{\cdot,j}
  \;=\; 
  \sum_{j=1}^k\frac{b_j}{c_j} X_{\cdot,j}.
\end{eqnarray}

\subsubsection{The detection statistic and its distribution}\label{sec:DetStatDistn}
We define the statistic
\begin{equation}\label{eqn:GLRstatistic}
  H_k 
  \;=\;
  \sum_{j=1}^k \frac{\Bigl(\sum_{i=1}^N \frac{x_{i,j}\,y_i}{\sigma_i^2}\Bigr)^2}{\sum_{i=1}^N \frac{x_{i,j}^2}{\sigma_i^2}}
  \;=\;
  2\times\log\biggl(\frac{p(y|\hat{\beta})}{p(y|\vec{0})}\biggr)
\end{equation}
(see also~(\ref{eqn:maxLrLimo})) which, under the null hypothesis of
the data~$y$ being purely noise, is $\chi^2$~distributed with
$k$~degrees of freedom.  Under the signal hypothesis, when a
signal~$s_{\beta^\star} = X \beta^\star$ is present in the data, the
corresponding figure \textsl{evaluated at the true parameter
values~$\beta^\star$},
\begin{equation}
  2\times \log\biggl(\frac{p(y|\beta^\star)}{p(y|\vec{0})}\biggr),
\end{equation}
will be Gaussian distributed 
with mean $\varrho^2$ 
and variance $4\varrho^2$, where
\begin{eqnarray}\label{eqn:SnrDefinition}
  \varrho^2 
  &=&
  \sum_{i=1}^N \frac{\bigl(\sum_{j=1}^k \beta^\star_j x_{i,j}\bigr)^2}{\sigma_i^2}
  \;=\;
  \sum_{i=1}^N \frac{(X\beta^\star)_i^2}{\expect\bigl[\epsilon_i^2\bigr]}
  \\
  &=&
  (X\beta^\star)^\prime \,\Sigma^{-1}\, (X\beta^\star)
\end{eqnarray}
is the true signal's \textsl{signal-to-noise ratio (SNR)}.
Consequently, for a signal of given SNR~$\varrho^2$, the expected
logarithmic likelihood ratio evaluated at the true parameters is
$\expect\Bigl[\log\bigl(\frac{p(y|\beta^\star)}{p(y|\vec{0})}\bigr)\Bigr]=\frac{1}{2}\varrho^2$,
while the likelihood ratio $\frac{p(y|\beta^\star)}{p(y|\vec{0})}$
follows a log\mbox{-}normal distribution with median
$\exp(\frac{1}{2}\varrho^2)$ and expectation
$\expect\Bigl[\frac{p(y|\beta^\star)}{p(y|\vec{0})}\Bigr]=\exp(\varrho^2)$.
The \textsl{maximized} likelihood ratio will be larger than that; the
statistic~$H_k$ follows a noncentral
$\chi^2_k(\varrho^2)$-distribution with noncentrality
parameter~$\varrho^2$, its expectation is $\varrho^2+k$ ,
so that
$\expect\Bigl[\log\bigl(\frac{p(y|\hat{\beta})}{p(y|\vec{0})}\bigr)\Bigr]=\frac{1}{2}\bigl(\varrho^2+k\bigr)$.
Note that the GW and signal-processing literature is sometimes
confusing, as both $\varrho^2$ and $H_k$, or their square roots, are
commonly referred to as the \textsl{SNR}.

In common signal detection problems, the signal model is usually only
\textsl{partially} linear, as suggested in Sec.~\ref{sec:WhyLiMo}, so
that analytical maximization over the ``linear'' parameters only
yields a maximized \textsl{conditional} likelihood, or profile
likelihood.  The statistic~$H_k$ then is proportional to the profile
likelihood, and (since the likelihood under the ``noise only'' null
hypothesis, $p(y|\vec{0})$, is a constant) constitutes a generalized
Neyman-Pearson test statistic.  This statistic, or its maximum over
additional parameters, is commonly referred to as a \textsl{detection
statistic}, as it is used to find the signal fitting the data best,
and to determine its significance.  The detection statistic's
distributions under null and alternative hypotheses as stated above
only apply for a single (conditional) likelihood maximization, i.e.,
for a given data set~$y$ and a given model matrix~$X$.  When
maximizing the profile likelihood over additional parameters (or
pieces of data), the testing problem turns into a \textsl{multiple
testing} problem, and the statistic's distribution will be an extreme
value statistic \cite{MGB,RoeverMessengerPrix2011}.  Since the
particular statistic~$H_k$ only comes up in the context of the
Gaussian model, we will in the following be mostly referring to the
more universal corresponding likelihood ratio
figure~$\frac{p(y|\hat{\beta})}{p(y|\vec{0})}=\exp(\frac{1}{2}H_k)$.

\subsection{Common implementation and terminology}\label{sec:commonImplementation}
In the GW data analysis literature, likelihoods and matched filters
are commonly expressed in terms of the \textsl{inner product} $\langle
a,b\rangle$ of real-valued functions (signal templates or data) $a$
and $b$, technically defined in terms of analytical Fourier
transforms,
\begin{eqnarray}\label{eqn:innerProductIntegral}
  \langle a,b \rangle 
  &=&
  \int_{-\infty}^{\infty} \frac{\tilde{a}(f)\,\tilde{b}(f)^\ast}{S_1(f)}\,\differential f
\end{eqnarray}
\cite{Finn1992,JaranowskiKrolakSchutz1998},
which in practice is implemented (analogously to the Whittle
likelihood) in terms of discrete Fourier transforms,
\begin{eqnarray}\label{eqn:innerProductSum}
  && \langle a,b \rangle 
  \\
  &=&
  2\sum_{j=0}^{N/2} \frac{\frac{\Delta_t}{N}
                \Bigl[\re\bigl(\tilde{a}(f_j)\bigr)\re\bigl(\tilde{b}(f_j)\bigr) 
                      \!+\! \im\bigl(\tilde{a}(f_j)\bigr)\im\bigl(\tilde{b}(f_j)\bigr)\Bigr]}{S_1(f_j)}.\nonumber
\end{eqnarray}
In terms of the linear models discussed in the previous section, this
is equivalent to a quadratic form
\begin{equation}
  \vec{a}^\prime \, \Sigma^{-1}\, \vec{b}
\end{equation}
as in Eqs.\ (\ref{eqn:bj}), (\ref{eqn:cj}) above.  Note that
especially in the context of the Student\mbox{-}$t$ model discussed
below, expression~(\ref{eqn:innerProductIntegral}) may be hard to
motivate, as it is continuous in frequency, but the corresponding
discrete expression (\ref{eqn:innerProductSum}) may readily be related
to expressions derived above.  In this terminology, the
signal-to-noise ratio of a signal~$s_\theta$ (\ref{eqn:SnrDefinition})
turns out as
\begin{eqnarray}\label{eqn:innerProductSnr}
  \varrho^2
  &=&
  \sum_j \frac{|\tilde{s}_\theta(f_j)|^2}{\frac{N}{4\Delta_t}S_1(f_j)}
  \;=\;
  2 \, \langle s_\theta,s_\theta \rangle,
\end{eqnarray}
the correlation of some data~$d$ with a template~$s_\theta$
(as in (\ref{eqn:bj})) simplifies to
\begin{eqnarray}\label{eqn:innerProductCorrelation}
  &&\sum_j \frac{\Bigl[\re\bigl(\tilde{d}(f_j)\bigr)\re\bigl(\tilde{s}_\theta(f_j)\bigr) 
                     + \im\bigl(\tilde{d}(f_j)\bigr)\im\bigl(\tilde{s}_\theta(f_j)\bigr)\Bigr]}{\frac{N}{4\Delta_t}S_1(f_j)}\nonumber \\
  &=&
  2 \,\langle d, s_\theta \rangle,
\end{eqnarray}
the likelihood ratio of some signal
template~$s$ for given data~$d$ is
\begin{eqnarray}\label{eqn:innerProductLikelihood}
  \log \Bigl(\frac{p(d|s_\theta)}{p(d|\vec{0})}\Bigr)
  &=&
  \frac{\sum_j \frac{|\tilde{d}(f_j)-\tilde{s}_\theta(f_j)|^2}{S_1(f_j)}}
       {\sum_j \frac{|\tilde{d}(f_j)|^2}{S_1(f_j)}}
  \\
  &=&
  2 \, \langle d,s_\theta \rangle - \langle s_\theta ,s_\theta \rangle
\end{eqnarray}
\cite{RoeverMeyerChristensen2011,Finn1992,JaranowskiKrolakSchutz1998},
and the \textsl{maximized} likelihood ratio for a signal that is a
linear combination of waveforms ($d=\sum_j\beta_j s_j+n$, see also
(\ref{eqn:maxLrLimo})) then is
\begin{eqnarray}\label{eqn:innerProductMaxLikelihood}
  \log \Bigl(\frac{p(d|\hat{\beta})}{p(d|\vec{0})}\Bigr)
  &=&
  \sum_j{}
  \frac{\langle d,s_{j} \rangle^2}{\langle s_{j} ,s_{j} \rangle  }.
\end{eqnarray}

An implementation of a matched filter in the GW context is concisely
described e.g.\ in \cite{AllenEtAl1999,AllenEtAl2005}.  The signal
searched for is a ``chirping'' binary inspiral waveform of increasing
frequency and amplitude, which is characterized by five parameters,
namely two mass parameters determining the phase/amplitude evolution,
and amplitude, phase and arrival time.  The signal waveform~$s$ for
given mass parameters~$\vartheta=(m_1,m_2)$ is (in analogy to
(\ref{eqn:sinusoidExample})) given in terms of ``sine'' and ``cosine''
components $s_{\mathrm{s},\vartheta}$ and $s_{\mathrm{c},\vartheta}$,
\begin{equation}\label{eqn:AllenEtAlFilter}
  s_\vartheta(t) \;=\; \beta_{\mathrm{s}} \, s_{\mathrm{s},\vartheta}(t-t_0) + \beta_{\mathrm{c}} \, s_{\mathrm{c},\vartheta}(t-t_0)
\end{equation}
\cite{AllenEtAl1999},
where $\beta_{\mathrm{s}}$ and $\beta_{\mathrm{c}}$ are determined by
the orbital phase and orientation of the binary system, and $t_0$
defines the signal arrival time.  The sine and cosine waveforms here
constitute the signal manifold's orthogonal ``basis vectors''.  The
actual matched-filter detection statistic is defined as
$\rho(t_0)=\sqrt{X^{\mathrm{s}} (t_0)^2 + X^{\mathrm{c}} (t_0)^2}$,
where
\begin{equation}\label{eqn:generalMatchedFilter}
  X^{\mathrm{s}/\mathrm{c}} (t_0) \;\propto\; \int \frac{\tilde{d}(f)\, (\tilde{s}_{\mathrm{s}/\mathrm{c},\vartheta}(f))^\ast\,\exp(-2\pi\imag f t_0)}{S_y(|f|)}\,\differential f
\end{equation}
\cite{AllenEtAl1999}, 
and where the exponential term does the time shifting of data and
template against each other.  For any given time shift~$t_0$, this
filter corresponds to (the square root of) the detection
statistic~$H_k$ above (\ref{eqn:GLRstatistic}).  Computing the matched
filter~(\ref{eqn:generalMatchedFilter}) across time points~$t_0$
yields the profile likelihood, the conditional likelihood (conditional
on time~$t_0$ and waveforms~$s_{\mathrm{s},\vartheta}$,
$s_{\mathrm{c},\vartheta}$) maximized over phase and amplitude. The
``overall'' maximum likelihood then is determined via a brute-force
search over~$t_0$ and over additional alternative signal waveforms
corresponding to different mass parameters~$\vartheta$. Note that the
search over arrival time~$t_0$ in~(\ref{eqn:generalMatchedFilter}) may
be efficiently implemented via another Fourier transform
\cite{AllenEtAl2005}. The matched-filtering algorithm is also
described in more detail in Appendix~\ref{sec:PseudocodeApp}.

In order to claim the \textsl{detection} of a signal, one needs to
determine a threshold for the detection statistic (the maximized
likelihood), with respect to some pre-specified false alarm rate.  The
detection statistic's distributions derived in
Sec.~\ref{sec:DetStatDistn} are likely not to be of much practical
relevance, due to common non-Gaussian or nonstationary features in the
data. Critical values for the detection statistic instead are commonly
computed using bootstrapping methods (see e.g.\
\cite{Brown2005,WasEtAl2010}).

\begin{figure*}
  \includegraphics[width=0.3\linewidth]{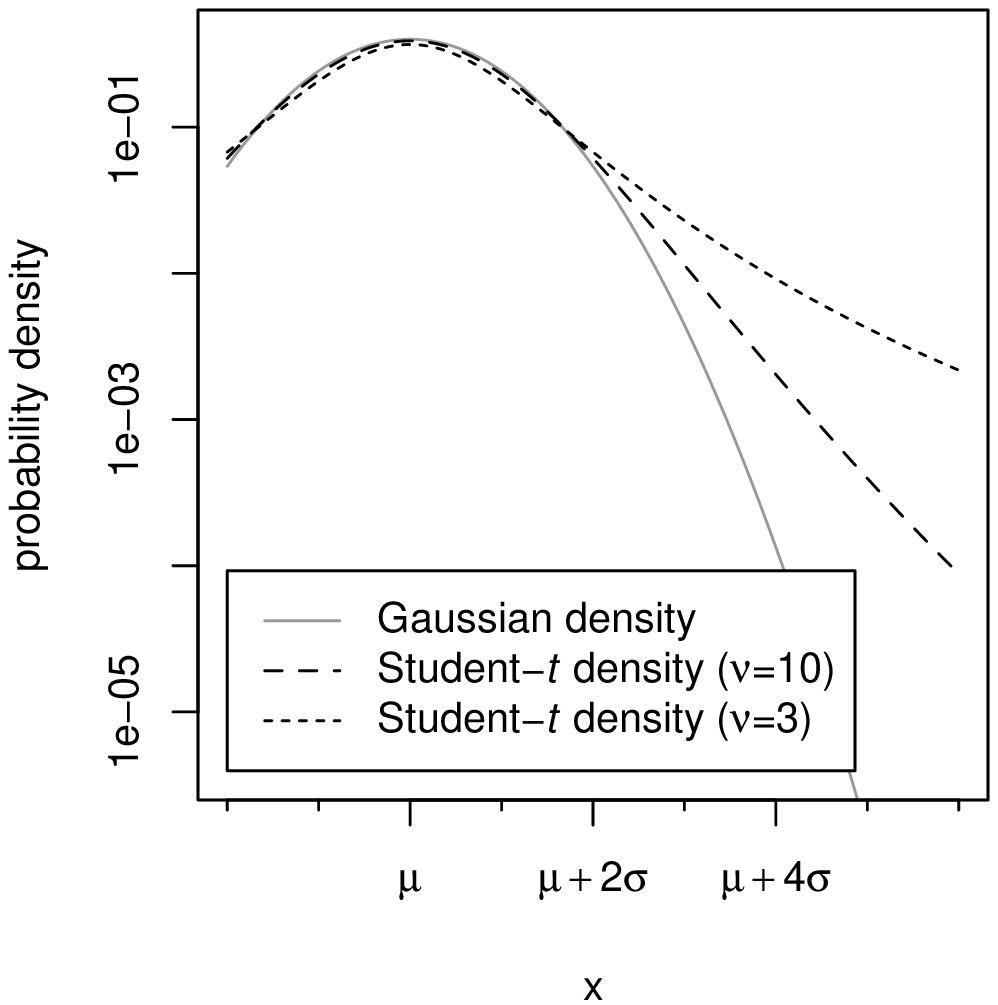}
  \hspace{0.1\linewidth}
  \includegraphics[width=0.3\linewidth]{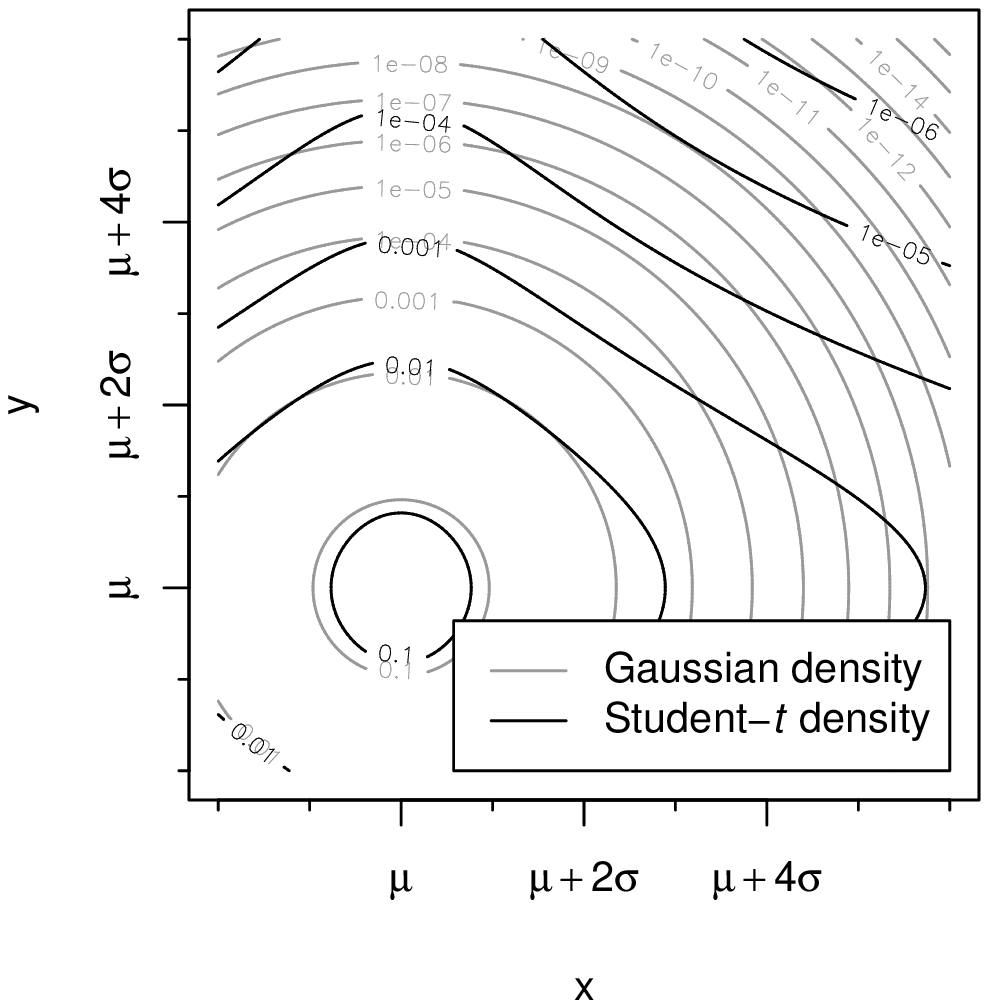}
  \caption{Density functions of Gaussian and Student\mbox{-}$t$
           distributions. The left panel shows univariate densities on
           the logarithmic scale.  The right panel shows density
           contours of the joint distribution of two independent
           Gaussian random variables in contrast with two independent
           Student\mbox{-}$t$ distributed variables of the same
           location~($\mu$) and scale~($\sigma$). The two
           Student\mbox{-}$t$ variables have differing
           degrees-of-freedom; the one corresponding to the
           $x$~axis has $\nu=3$, while the one along the
           $y$~axis has $\nu=10$. }
  \label{fig:densities}
\end{figure*}

\section{The Student-t filter}\label{sec:StudentTFilter}
\subsection{The Student-t noise model}\label{sec:StudentTModel}
The Student\mbox{-}$t$ model for time series analysis was introduced
in \cite{RoeverMeyerChristensen2011} as a generalization of the
commonly used Gaussian model described in the previous section.  The
Student\mbox{-}$t$ distribution has an additional
\textsl{degrees-of-freedom} parameter, essentially controlling the
distribution's heavy-tailedness, i.e., the allowance for large
outliers.  The Student\mbox{-}$t$ likelihood function is given by
\begin{eqnarray}
  && p\bigl(\vec{d} | \theta \bigr)\nonumber \\
  &\propto& 
  \!\prod_{j} \biggl(1 + \frac{1}{\nu_j}\,\frac{\bigl|\tilde{d}(f_j)-\tilde{s}_\theta(f_j)\bigr|^2}{\frac{N}{4\Delta_t}\,S_1(f_j)}\biggr)^{-\frac{\nu_j + 2}{2}}\label{eqn:StudentTLikeli1}
  \\
  &=& 
  \!\exp\biggl(- \! \sum_{j} {\textstyle \frac{\nu_j + 2}{2}} \log\biggl[1 \!+\! \frac{1}{\nu_j}\frac{\bigl|\tilde{d}(f_j)\!-\!\tilde{s}_\theta(f_j)\bigr|^2}{\frac{N}{4\Delta_t}\,S_1(f_j)}\biggr]\biggr)\label{eqn:StudentTLikeli2}
\end{eqnarray}
\cite{RoeverMeyerChristensen2011}.  According to this model, the
residuals ($\re(\tilde{n}(f_j))$, $\im(\tilde{n}(f_j))$) within each
Fourier frequency bin~$j$ follow a \textsl{bivariate
Student\mbox{-}$t$ distribution} \cite{BDA}
with location~$\mu=\vec{0}$,
scale matrix~$\Sigma=\frac{N}{4\Delta_t}\left(\begin{array}{cc} S_1(f_j) & 0 \\ 0 & S_1(f_j) \end{array}\right)$, 
degrees-of-freedom~$\nu_j>0$ and implicit dimension~$2$\@. 
This implies that 
(i)~residuals in different frequency bins are independent;
(ii)~ residuals within the same bin are uncorrelated, but dependent; and
(iii)~the marginal distribution of each individual residual is a Student\mbox{-}$t$ distribution with scale proportional to~$S_1(f_j)$ and degrees-of-freedom~$\nu_j$.
Decreasing values of the degrees-of-freedom parameters~$\nu_j$ imply a
heavier-tailed distribution, and in the limit of
$\nu_j\rightarrow\infty$ the model again reduces to the Gaussian
model.

Besides simply constituting a heavier-tailed noise model, the
Student\mbox{-}$t$ model arises as a generalization of the Gaussian
model when the power spectral density $S(f_j)$ is treated as
uncertain, where the degrees-of-freedom parameter~$\nu_j$ denotes the
(prior) precision \cite{RoeverMeyerChristensen2011}.  So the model is
not only applicable in contexts where the noise itself is in fact
$t$~distributed, but also in cases where it is Gaussian, but
the noise spectrum is a~priori only known to a certain accuracy.
Alternatively, the same model would result when the noise spectrum
itself was assumed to be randomly deviating from the scale parameter
$S_1(f)$, according to a $\chi^2$~distribution, e.g.\ because it is
only estimated with some uncertainty, which in fact resembles the
original motivation for introducing Student's $t$\mbox{-}distribution
in the context of the $t$\mbox{-}test and related procedures
\cite{Gosset1908,MGB}.  Both randomness or uncertainty of the noise
PSD technically lead to the same likelihood expression here
\cite{RoeverMeyerChristensen2011}.
In general, the interpretation of the scale parameter~$S_1(f)$ in the
contexts of the Gaussian and the Student\mbox{-}$t$ model is not
necessarily exactly the same.  For the Gaussian model, it may be
defined via the expected power $S_1(f_j) =
\expect\bigl[2\frac{\Delta_t}{N}\,|\tilde{n}(f_j)|^2\bigr]$, while for
the Student\mbox{-}$t$ model this only holds in the limiting case of
great certainty ($\nu\rightarrow\infty$).
Within the Student\mbox{-}$t$ model, the $S(f)$ term specifies the
scale of the uncertain PSD parameter and the expected power is in fact
given by 
$\expect\bigl[2\frac{\Delta_t}{N}\,|\tilde{n}(f_j)|^2\bigr] = \frac{\nu_j}{\nu_j-2}\,S_1(f_j)$.
The choice of the degrees-of-freedom parameter~$\nu_j$ as well as the
spectrum parameter~$S_1(f_j)$ may be approached in different ways and
may for the filtering purpose eventually be considered a matter of
tuning \cite{RoeverMeyerChristensen2011}.
In the example in Sec.~\ref{sec:example} below, we simply kept the
scale parameter~$S_1(f_j)$ to be the estimated noise spectrum as in
the Gaussian case, and fitted a common degrees-of-freedom
parameter~$\nu_j=\nu$ for all frequency bins to the empirical data.

\subsection{Comparison to the Gaussian model}
When comparing to the Gaussian distribution, first of all the
Student\mbox{-}$t$ distribution exhibits \textsl{heavier tails}, i.e.,
the probability for obtaining ``large'' values (relative to the
distribution's scale) is much greater.  While the density functions
are very similar within the range of $\mu\pm 2\sigma$, where the bulk
of probability is concentrated, the densities' ratio will grow
indefinitely toward the distributions' tails (see
Fig.~\ref{fig:densities}). The degrees-of-freedom parameter~$\nu$
controls the distribution's heavy-tailedness; a setting of $\nu=1$
yields the ``pathological'' Cauchy distribution, for $\nu>2$ the
variance is finite, and in the limit of $\nu\rightarrow\infty$ it again
approaches the Gaussian distribution.

Another discriminating feature is the shape of the density contours.
While a Gaussian density will always have elliptical contours, the
Student\mbox{-}$t$ distribution is different in that its contours are
rather diamond-shaped, with elongations pointing along the principal
axes (see Fig.~\ref{fig:densities}).  This way the Student\mbox{-}$t$
model does not only allow for larger outliers, but it also considers
outliers more likely to occur only in individual variables rather than
jointly in all variables. Note that this latter effect follows from
the fact the different frequency bins are \textsl{stochastically
independent} and not merely \textsl{uncorrelated}
\cite{KelejianPrucha1985,BreuschRobertsonWelch1997}.
Since the two (real and imaginary) residuals within each Fourier
frequency bin follow a joint, bivariate, $t$~distribution, the
density contours \textsl{within} bins will still be
spherical---otherwise a strange phase/amplitude dependence would be
implied for the Fourier-domain model. The effect of independent
Student\mbox{-}$t$ variables only comes to bear \textsl{between}
frequency bins.

An important difference to note between the Gaussian and
Student\mbox{-}$t$ model is that the least-squares fitting that
results from the Gaussian model will actually be a ML procedure for
any model within the wider class of ``elliptically symmetric'' models
for the noise residuals (including e.g.\ a Student\mbox{-}$t$ model
with merely \textsl{uncorrelated}, but not \textsl{independent}
residuals) \cite{KelejianPrucha1985,BreuschRobertsonWelch1997}. The
Student\mbox{-}$t$ model described here hence advances into a
fundamentally different class of models.

Student\mbox{-}$t$ or similar models are commonly used in parameter
estimation contexts as robust alternatives to the Gaussian model that
are less sensitive to outliers in the data
\cite{LangeEtAl1989,Geweke1993,Divgi1990,McDonaldNewey1988}.  Such
models may be motivated in a ``top-down'' manner by the observation
that the data do not actually fit the Gaussianity assumption, or also
in a ``bottom-up'' way by pointing out that the resulting least-squares
procedures are very sensitive to occasional outliers in the data.  In
the spirit of the latter viewpoint, the concept of
\textsl{M~estimation} was introduced, which aims at ``fixing''
outlier-sensitive least-squares procedures by replacing them by more
robust statistics corresponding to more favorable \textsl{influence
functions} \cite{Hampel,Huber}.  Similar approaches, namely
down-weighting or ignorance of outliers in the data, have been
proposed in the context of gravitational-wave detection before
\cite{Creighton1999,AllenEtAl2002}, and the Student\mbox{-}$t$
assumption may in fact be considered a special case of
M~estimation \cite{Divgi1990,McDonaldNewey1988}.

Another fix that is commonly applied in GW data analysis is the
\textsl{$\chi^2$~veto} \cite{Allen2005}, which is a figure computed
along with a detection statistic that is supposed to discriminate
actual signals from noise bursts. Such noise events may show little
similarity with the signal template, but may often, due to
non-negligible correlation with the template and very large power,
still seem to indicate the presence of a signal. The $\chi^2$~veto
then essentially checks for excess power that is inconsistent with the
shape of the signals aimed for and that way will rule out such alleged
detections.
The consideration of excess residual power is also implicitly
happening in the Student\mbox{-}$t$ model.  From the different
likelihood formulations ((\ref{eqn:GaussianLikelihood}),
(\ref{eqn:StudentTLikeli2})) one can write down the corresponding
likelihood ratios for some data~$d$ and a signal template~$s_\theta$,
\begin{eqnarray}
  &&\log\biggl(\frac{p(d|\theta,\mbox{Gauss})}{p(d|\vec{0},\mbox{Gauss})}\biggr)\nonumber\\
  &=&
  \sum_j \frac{1}{2}\left(\frac{\bigl|\tilde{d}(f_j)\bigr|^2}{\frac{N}{4\Delta_t}\,S_1(f_j)}
                -\frac{\bigl|\tilde{d}(f_j)-\tilde{s}_\theta(f_j)\bigr|^2}{\frac{N}{4\Delta_t}\,S_1(f_j)}\right),
  \label{eqn:GaussianLR}
\end{eqnarray}
\begin{eqnarray}
  &&\log\biggl(\frac{p(d|\theta,\mbox{Student})}{p(d|\vec{0},\mbox{Student})}\biggr)\nonumber\\
  &=&
  \sum_j {\textstyle\frac{\nu_j+2}{2}} \log\left(\frac{1+\frac{1}{\nu_j}\frac{\bigl|\tilde{d}(f_j)\bigr|^2}{\frac{N}{4\Delta_t}\,S_1(f_j)}}{1+\frac{1}{\nu_j}\frac{\bigl|\tilde{d}(f_j)-\tilde{s}_\theta(f_j)\bigr|^2}{\frac{N}{4\Delta_t}\,S_1(f_j)}}\right).
  \label{eqn:StudentLR}
\end{eqnarray}
In both of the above cases the likelihood ratio is a function of 
the ``data power''~$\frac{\bigl|\tilde{d}(f_j)\bigr|^2}{\frac{N}{4\Delta_t}\,S_1(f_j)}$
and the ``residual power''~$\frac{\bigl|\tilde{d}(f_j)-\tilde{s}_\theta(f_j)\bigr|^2}{\frac{N}{4\Delta_t}\,S_1(f_j)}$,
i.e., the data's normalized sum-of-squares in each frequency bin~$j$
before and after subtracting the signal~$s_\theta$.
For the Gaussian case, a ``data power'' of~10 and a ``residual power''
of~1 in the $j$th bin would have the same effect on the likelihood
ratio as if the numbers were, say, 1010 and 1001 instead; the only
relevant figure is their difference.
In the Student\mbox{-}$t$ model, the latter case would lead to a lower
likelihood ratio; here not only the amount by which the
signal~$s_\theta$ is able to reduce the sum-of-squares is relevant,
but so is its magnitude relative to the remaining residual term.  The
additional feature of the ML fit that is intrinsically considered in
the Student\mbox{-}$t$ likelihood ratio (\ref{eqn:StudentLR}) is
essentially the corresponding \textsl{coefficient of determination}
($R^2$) \cite{NeterEtAl}.
As will become obvious in the following, when the actual
implementation is described, the generalization to the
Student\mbox{-}$t$ model will on the technical side essentially
replace the least-squares procedure by an adaptive version. The
adaptation step again ensures that excess residual noise power will
downweight the supposed significance of a signal.

\subsection{Likelihood maximization: the EM-algorithm}\label{sec:StudentLikeliMaxi}
While likelihood maximization in the Gaussian model boils down to
least-squares fitting, the maximization step is not quite as simple
for the Student\mbox{-}$t$ model. However, due to the structure of the
problem, the \textsl{expectation-maximization (EM) algorithm} may be
used to efficiently maximize the likelihood function
\cite{BDA,DempsterLairdRubin1977}.  In order to apply the
EM~algorithm, the likelihood expression needs to be reformulated. The
Student\mbox{-}$t$ likelihood may be viewed as a \textsl{marginal}
likelihood, averaging out a set of unknown variance
parameters~$\vec{\sigma}^2$ \cite{RoeverMeyerChristensen2011}. Each of
the variance parameters~$\sigma_j^2$ then corresponds to the power
spectral density at the $j$th Fourier frequency bin.  The EM
algorithm's details as applied to the present problem are derived in
detail in Appendix~\ref{sec:EMApp} below.  It turns out that
maximization of the Student\mbox{-}$t$ likelihood may be done in an
iterative manner, where each iteration again requires a weighted
least-squares fit as in the Gaussian matched filter.  The EM~algorithm
requires a starting value~$\theta_0$ for the signal parameters.
Given~$\theta_0$, the expression
\begin{equation}
  \mathcal{E}(\theta_0, \theta)
  =
  -{\textstyle\frac{1}{2}}\sum_j {\textstyle \frac{\bigl|\tilde{d}(f_j)-\tilde{s}_\theta(f_j)\bigr|^2}{\frac{N}{4\Delta_t} \bigl(\frac{\nu_j}{\nu_j+2}S_1(f_j)+\frac{2}{\nu_j+2}\frac{2\Delta_t}{N}\bigl|\tilde{d}(f_j)-\tilde{s}_{\theta_0}(f_j)\bigr|^2\bigr)}}
  \label{eqn:EM_expectation2}
\end{equation}
is maximized with respect to the parameter vector~$\theta$.  The
parameter value maximizing the above expression then constitutes the
new~$\theta_0$ value, for which the expression again is maximized, and
so forth. The resulting sequence of parameter values then converges to
the maximum likelihood estimate~$\hat{\theta}$ \cite{BDA}.

Maximizing the above expression (\ref{eqn:EM_expectation2}) again
amounts to a weighted least-squares fit, exactly as in the case of the
Gaussian matched filter (see also the corresponding likelihood
expression~(\ref{eqn:GaussianLikelihood})).  The Student\mbox{-}$t$
filter will therefore generalize the Gaussian matched filter by
replacing the least-squares procedure by an iterative, or adaptive,
least-squares fit.
Note that the denominator in~(\ref{eqn:EM_expectation2}) simply is a
weighted average of the noise spectrum (as
in~(\ref{eqn:GaussianLikelihood})) and the previous iteration's
residual noise power, where the degrees-of-freedom parameter~$\nu$
defines the relative weighting.
Instead of the ``plain'' weighted least-squares match that is done in
the Gaussian filter, the EM-iterations adapt the weights (the
denominator in (\ref{eqn:EM_expectation2}), which in the Gaussian
model was the a~priori known, fixed noise spectrum) to the residual
noise power as found in the data, and the level of adaptation is
regulated by the degrees-of-freedom parameter~$\nu$.

The (ML) detection statistic does not follow a simple distributional
form as in the Gaussian model, but in the example below one can
already see that both statistics still behave similarly. The
generalized likelihood ratio statistic will, by Wilks' theorem, in
fact still approximately follow a $\chi^2$~dis\-tri\-bu\-tion
\cite{MGB,Wilks1938}.

\subsection{The filter implementation}
\label{sec:StudentFilterImplementation}
As for the Gaussian matched filter, the aim again is to maximize the
likelihood (\ref{eqn:StudentTLikeli2}), i.e., find best-fitting
parameter values~$\hat{\theta}$ in parameter space.  Again, it is
advantageous if the signal model can (at least partly) be formulated
as a linear model.

There are two obvious points in the matched-filtering procedure at
which one could insert the EM-iterations in order to generalize it to
the Student\mbox{-}$t$ case: either at the level of each (originally
analytical) maximization over linear model coefficients (usually
corresponding to amplitude and phase), or at a higher level, iterating
over linear coefficients as well as the signal arrival time
parameter. It is not obvious whether one implementation is more
sensitive than the other, but there definitely are differences in the
implied computational costs. Both approaches are described and
discussed in more detail in Appendix~\ref{sec:PseudocodeApp}.
An implementation of the latter algorithm, together with the analogous
matched filter, is available in \cite{bspec2011}.
In case of a brute-force search over additional signal parameters
(i.e., a ``template bank''), one could in fact consider moving the
EM-level yet another stage higher.

As a starting parameter value~($\theta_0$) for the algorithm, one
could, for example, use the null vector or an initial least-squares
fit. As a stopping criterion, one could terminate the algorithm once
the improvement in loga\-rith\-mic likelihood from the previous
iteration falls below some threshold, or when some maximum number of
iterations is reached.  Note that---unlike for the Gaussian linear
least-squares fit---the (conditional) likelihood might actually be
multimodal \cite{MakelainenEtAl1981}, so that different starting
values might lead to different results. It is not obvious whether this
occurs frequently in practice, or rather requires particularly rare
pathological circumstances; however, it does not appear to pose a
problem in the example below.

\begin{figure*}
  \includegraphics[width=0.3\linewidth]{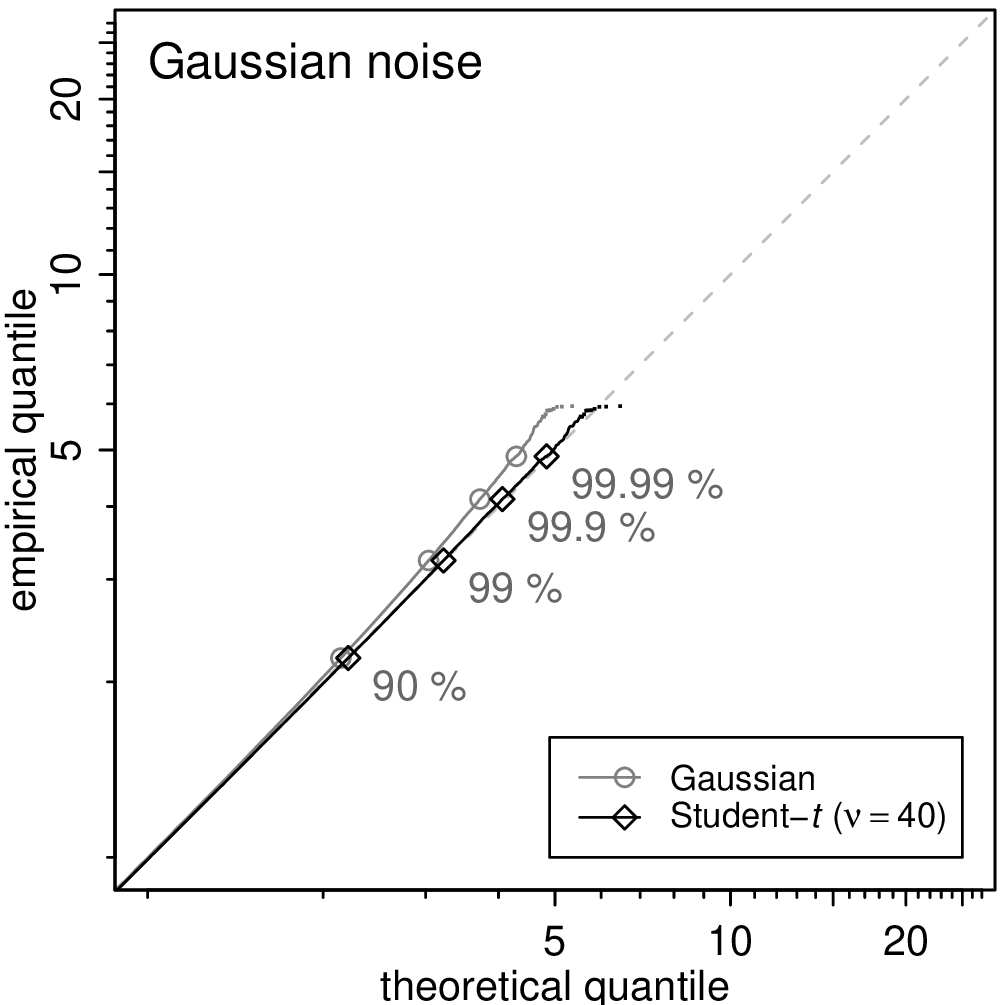}
  \hspace{0.1\linewidth}
  \includegraphics[width=0.3\linewidth]{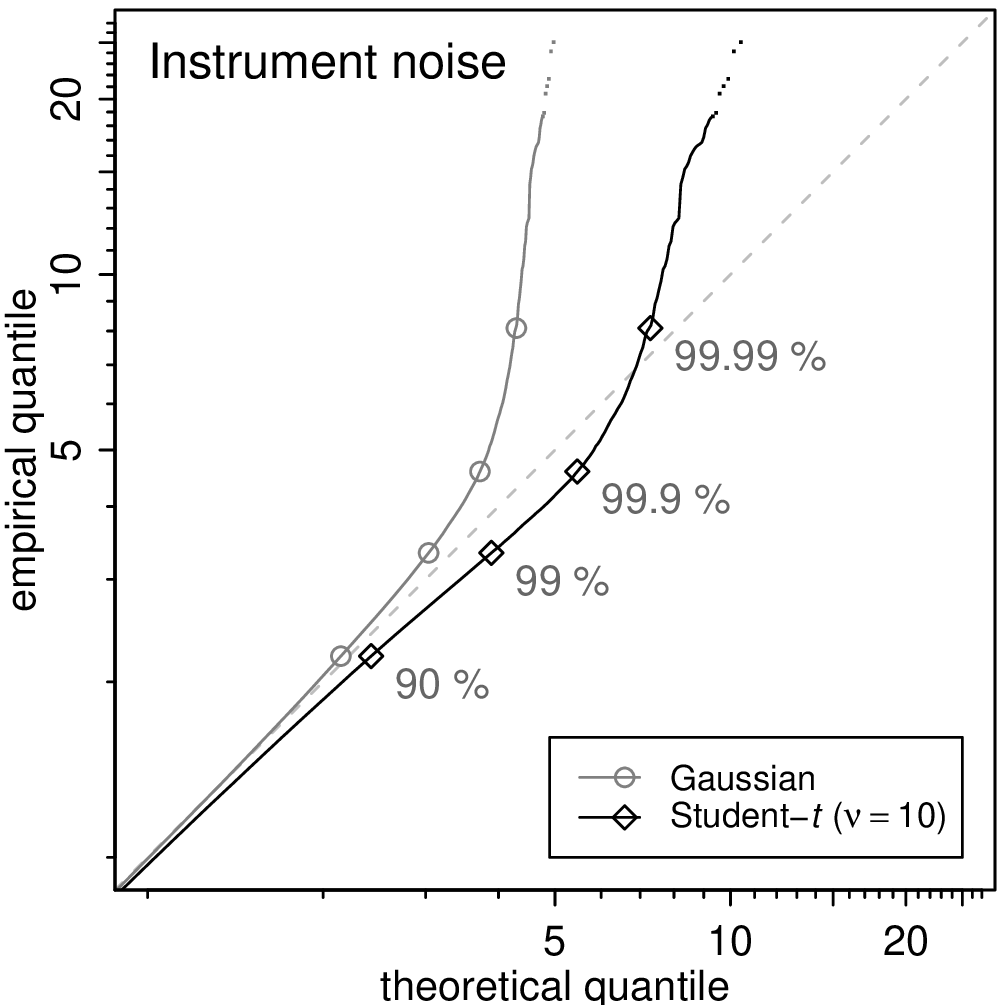}
  \caption{Quantile-quantile~plots (Q\mbox{-}Q~plots) of the
           empirically found normalized residual noise power
           (\ref{eqn:empiricalNoisePower}) versus its theoretical
           values assuming Gaussian and Student\mbox{-}$t$ models.
           The marks indicate particular quantiles corresponding to
           powers of~10 in tail probability. The 10~largest empirical
           samples are shown as individual dots; the remaining
           quantiles are connected by a line.}
  \label{fig:qqplots}
\end{figure*}

\section{Filtering experiment on actual data}\label{sec:example}
\subsection{General}
Besides any theoretical or heuristic arguments why a
Student\mbox{-}$t$ based filter may improve detection, the figure of
eventual relevance is going to be the resulting improvement in
detection efficiency when applied to actual data --- keeping in mind
the additional complication and computational cost.
In the following, we will demonstrate the filter's performance in a
minimalistic, yet realistic toy problem.
To that end, we will set up a filter for a certain kind of
parametrized signal, and then test it against a conventional matched
filter using injections of simulated signals.  For the additive noise,
we will use both simulated Gaussian noise as well as actual
gravitational-wave detector instrument noise.
Detection efficiency is going to be measured via the \textsl{receiver
operating characteristic (ROC)} curve, allowing one to compare detection
probabilities for given false alarm probabilities, or vice versa.

In order to make the example realistic, we require a nontrivial signal
waveform to be searched for; in particular the waveform should not be
monochromatic, but should instead span a wider range of Fourier
frequencies. There should be parameters to be maximized over
analytically as well as numerically, and we should use noise that is
non-Gaussian or nonstationary.
The example described in the following mimics the setup of a search
for binary inspiral signals in interferometric gravitational-wave
detector data (see e.g.~\cite{AllenEtAl1999}). The noise data are
taken from an actual detector, and, for comparison, a second data set
of simulated, Gaussian noise of a realistic noise spectrum is used in
parallel. The ``search'' being performed however is much simplified
and not intended to be exhaustive or to span an astrophysically
sensible parameter range.

\subsection{The data}
The data used in the following examples are going to be either
simulated Gaussian noise with a power spectral density corresponding
to LIGO's initial design sensitivity
\cite{DamourIyerSathyaprakash2001}, or real instrument noise from
LIGO's Livingston interferometer, taken during LIGO's fifth science run
(``S5'') in late~2005 \cite{AbbottEtAl2009a}.
The data will be considered in chunks of 8~seconds length, downsampled
to a sampling rate of $1024\,\mbox{Hz}$,
and windowed using a Tukey window tapering 10\% of the data 
(5\% at each end).
The noise's power spectral density~$S_1(f)$ is estimated essentially
using Welch's method \cite{Welch1967}, by considering the empirical
power in the 32~preceding data segments, and taking the median as a
robust estimator.  The figures shown in the following are each based
on $100\,000$ such data chunks.

The signal waveform searched for here is taken to be a binary inspiral
waveform approximated to the 2.0 post-Newtonian order
\cite{TanakaTagoshi2000}. The same waveform family is used for both
injections as well as in the detection stage, and it has five free
parameters: chirp mass ($m_c$), mass ratio ($\eta$), coalescence time
($t_c$), coalescence phase ($\phi_c$), and amplitude ($A$).
The signal waveforms injected into the data were all done at the same
mass parameters ($m_c=4.5$, $\eta=0.25$),
and the amplitude is set such that the signal's SNR (as computed based
on the current PSD estimate) is
$\varrho=\sqrt{\varrho^2}=5.257$
so that
$\expect\Bigl[\log \bigl(\frac{p(y|\beta^\star)}{p(y|\vec{0})}\bigr)\Bigr]=\frac{1}{2}\varrho^2 = \log(10^6)$
and 
$\expect\Bigl[\frac{p(y|\beta^\star)}{p(y|\vec{0})}\Bigr]=\exp\bigl(\varrho^2\bigr) = 10^{12}$
(see also Sec.~\ref{sec:DetStatDistn}).
Each 8\mbox{-}second chunk of data is eventually analyzed twice, 
with and without a signal injection.

\subsection{Setting the degrees-of-freedom parameter}\label{sec:DFSetting}
In order to determine a suitable degrees-of-freedom parameter~$\nu$
for the Student\mbox{-}$t$ model, we considered the tail behavior of
the noise.  If the Gaussian (Whittle) model was accurate, then the
normalized Fourier-domain noise power at the $j$th frequency bin,
\begin{equation}\label{eqn:empiricalNoisePower}
  \frac{\bigl|\tilde{n}(f_j)\bigr|}{\sqrt{\frac{N}{4\Delta_t}S_1(f_j)}},
\end{equation}
being the square root of the sum of two independent standard Gaussian
random variables (see Sec.~\ref{sec:GaussModel}), should follow a
\textsl{Rayleigh distribution}.  The residuals' normalization here is
done --- in analogy to the computations done in an actual search ---
via the \textsl{estimated} noise spectrum, as described in the
previous subsection.  We are only considering the binned noise
\textsl{power} here (and not the individual real and imaginary
components) as this is the relevant figure entering both the Gaussian
as well as the Student\mbox{-}$t$ likelihoods
((\ref{eqn:GaussianLikelihood}), (\ref{eqn:StudentTLikeli2}),
(\ref{eqn:GaussianLR}), (\ref{eqn:StudentLR})).  Under the
Student\mbox{-}$t$ model, instead of being Rayleigh distributed, the
power (\ref{eqn:empiricalNoisePower}) would instead follow a similar,
more heavy-tailed distribution. We will refer to the
Student\mbox{-}$t$ power's distribution as the
``\textsl{Student-Rayleigh}'' distribution here; more details on this
distribution's particular form are given in
Appendix~\ref{sec:StudentRayleighApp}.

\begin{figure*}
  \includegraphics[width=0.3\linewidth]{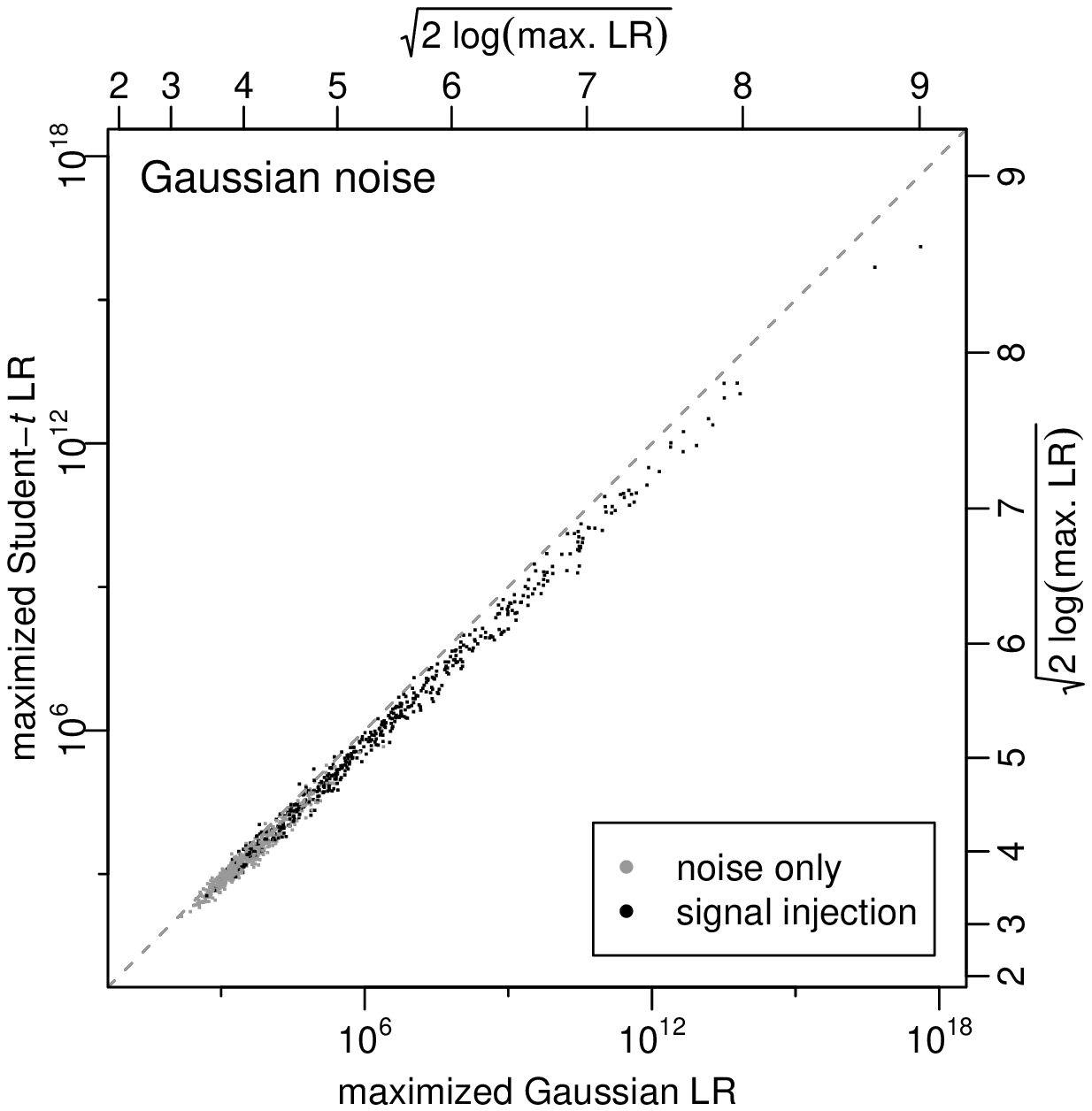}
  \hspace{0.1\linewidth}
  \includegraphics[width=0.3\linewidth]{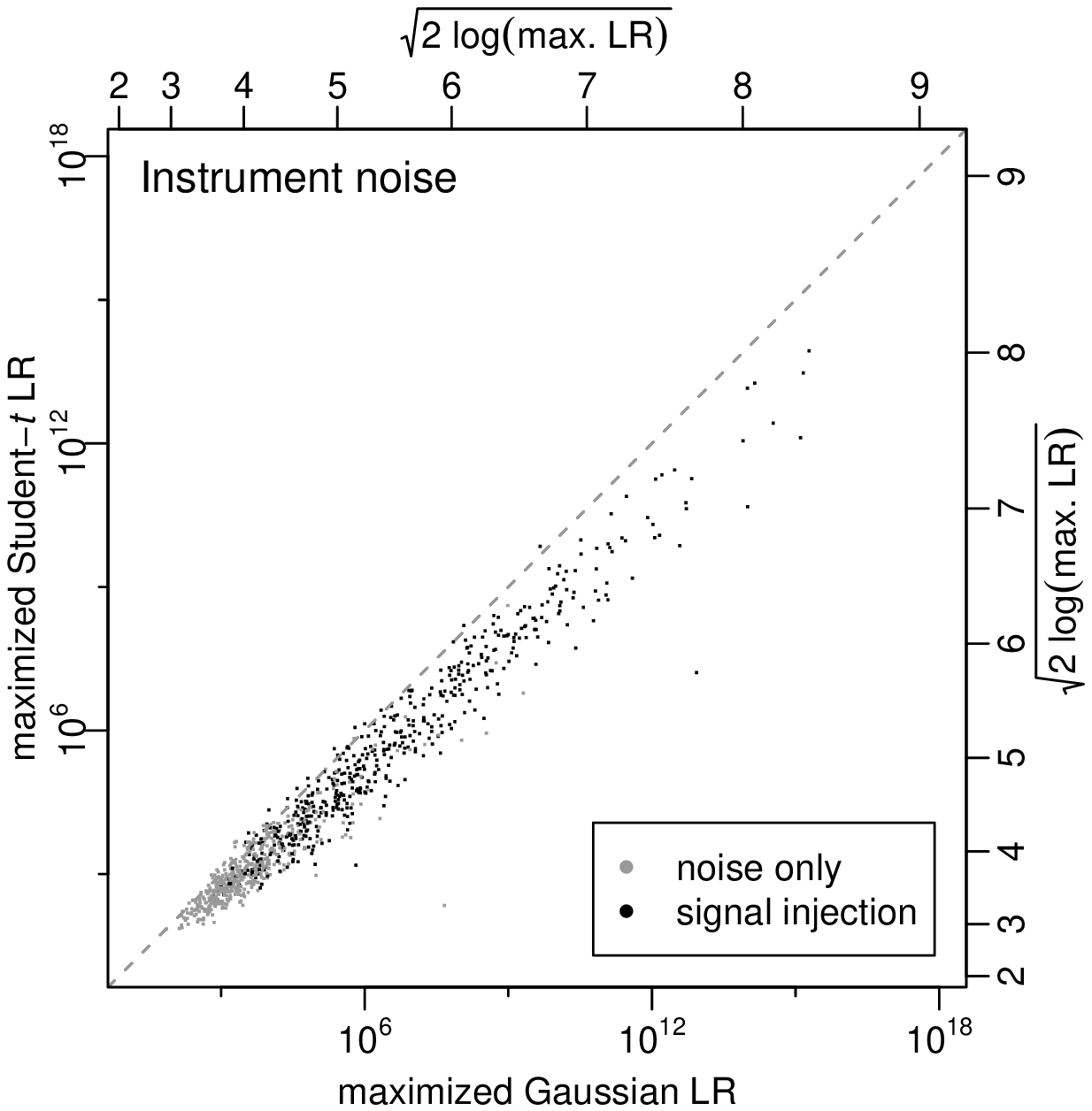}
  \caption{Detection statistics (maximized likelihood ratios) based on
           Gaussian and Student\mbox{-}$t$ models for simulated
           Gaussian data (left panel) and actual interferometer noise
           (right panel).  Injections were of SNR $\varrho=5.257$.}
  \label{fig:detectionStats}
\end{figure*}

We investigated the empirical distribution of actual noise residuals,
for both simulated and actual instrumental data. For the simulated
data, this will account for effects of finite sample size, windowing
and PSD estimation, and for actual data it will in addition give some
insight into the effects of realistic nonstationarities or
non-Gaussianities in actual measurement noise.  The noise samples are
based on the residuals from 200 eight-second noise realizations of
either simulated Gaussian noise, or actual instrument noise from
LIGO's Livingston interferometer. The residuals
(\ref{eqn:empiricalNoisePower}) are each normalized via a PSD estimate
from 32~preceding noise samples, as described in the previous section,
yielding a total of $800\,000$ residuals. The data used here did not
overlap with the data used in the following detection experiment.

Figure~\ref{fig:qqplots} shows quantile-quantile plots
(Q\mbox{-}Q~plots) illustrating how well the models fit the actual
data.  The axes indicate theoretical (Rayleigh or Student-Rayleigh)
quantiles, and the empirical quantiles as found in the data.  If a
model fits the data well, both theoretical and empirical quantiles
should coincide, so that the quantiles follow a straight, diagonal
line.
%
%
A mismatch between model and data results in a differently shaped
curve; in particular, if the data are more heavy-tailed than predicted
by the model, the curve will show an upward bend
\cite{WilkGnanadesikan1968}.

One can see that the actual data exhibit heavier tails in both cases
of simulated, Gaussian noise as well as the instrument noise.  In the
case of Gaussian noise this is due to the estimation uncertainty in
the noise spectrum. If we had been using the mean instead of the
median to estimate the noise PSD, then the distribution of normalized
noise residuals should be \textsl{exactly} Student\mbox{-}$t$ with
degrees of freedom equal to twice the number of noise samples averaged
over (here,~$32\!\times\! 2=64$) \cite{Gosset1908,MGB}.  For the
median estimation method, this is only approximately true, but
apparently still roughly accurate; a maximum-likelihood fit for~$\nu$
suggests a value of $\nu\approx 40$ here.  For the case of Gaussian
data, the mismatch between assumed and observed quantiles is minimal
anyway.

For the real interferometer noise, the discrepancy between Gaussian
model and actual data is more dramatic; in the distribution's tails,
the empirical quantiles are significantly larger than the assumed
quantiles.
For example, according to the Gaussian model, $99.99\,\%$ of the
samples should be $\leq 4.3$, while empirically the $99.99\,\%$
quantile lies at $8.1$ for actual instrument noise (see the right
panel of Fig.~\ref{fig:qqplots}).
A Student\mbox{-}$t$ model seems to fit the data better, especially in
the distributions' tails, although discrepancies in the extreme
outliers are still large. Trying to estimate the degrees-of-freedom
parameter~$\nu$ from different subsets of the empirical data yields ML
estimates roughly in the range from~5 to~50; in the following we
simply fixed the parameter at~$\nu=10$ for the simulations involving
actual data.  A value of $>40$ would not seem to make sense here (even
if the data were perfectly Gaussian) and in the simulation results
below we found that detection performance seemed to depend only weakly
on~$\nu$ as long as it was roughly in the range 5--20.  While the
Student\mbox{-}$t$ distribution does not fit the data perfectly, it
seems to fit better than the Gaussian model.  Instead of only fitting
the degrees-of-freedom parameter, one could actually in addition also adapt the
$t$~distribution's scale to the data (see also
Sec.~\ref{sec:StudentTModel}, or \cite{RoeverMeyerChristensen2011}).

\begin{figure*}
  \includegraphics[width=0.3\linewidth]{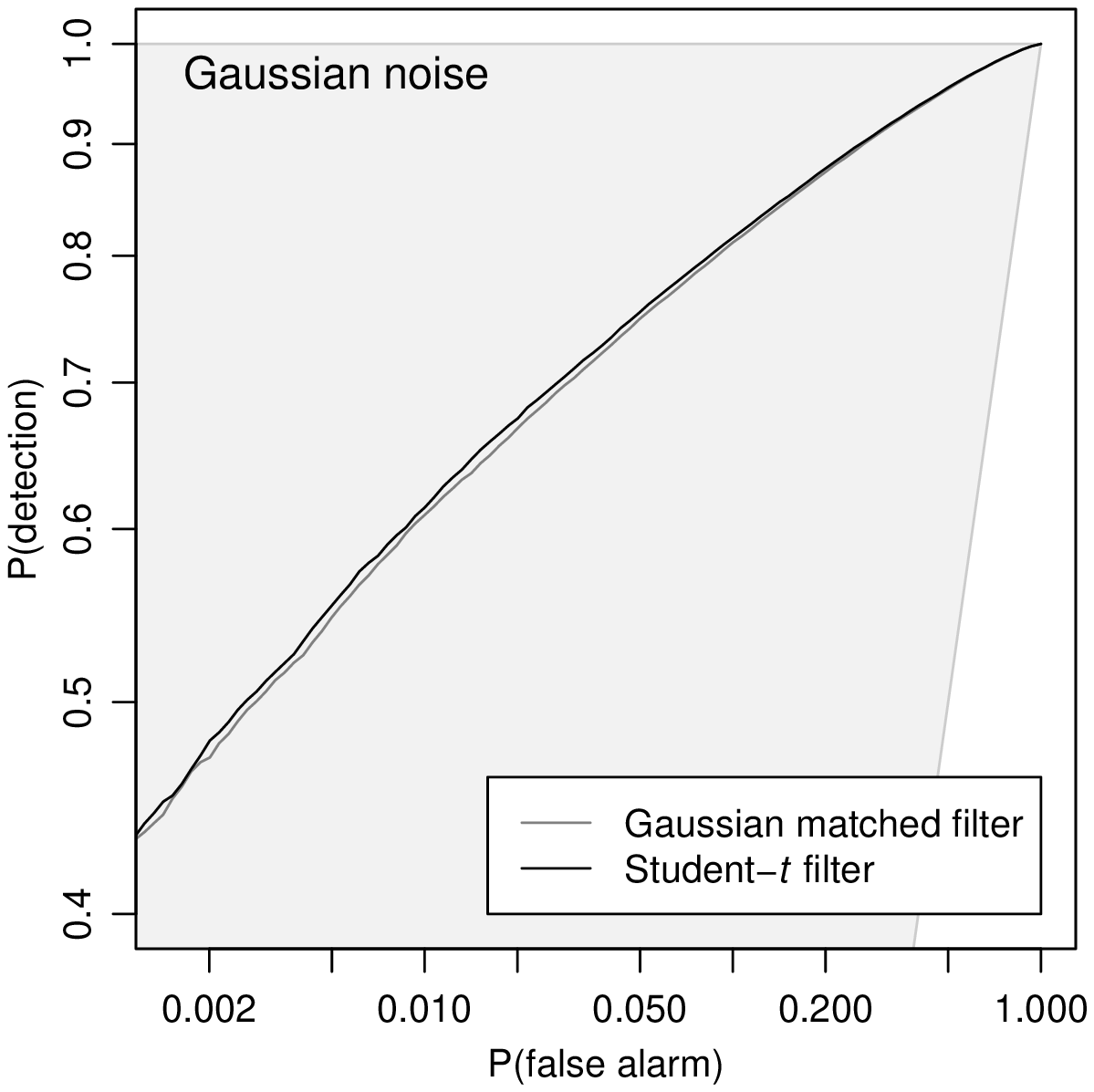}
  \hspace{0.1\linewidth}
  \includegraphics[width=0.3\linewidth]{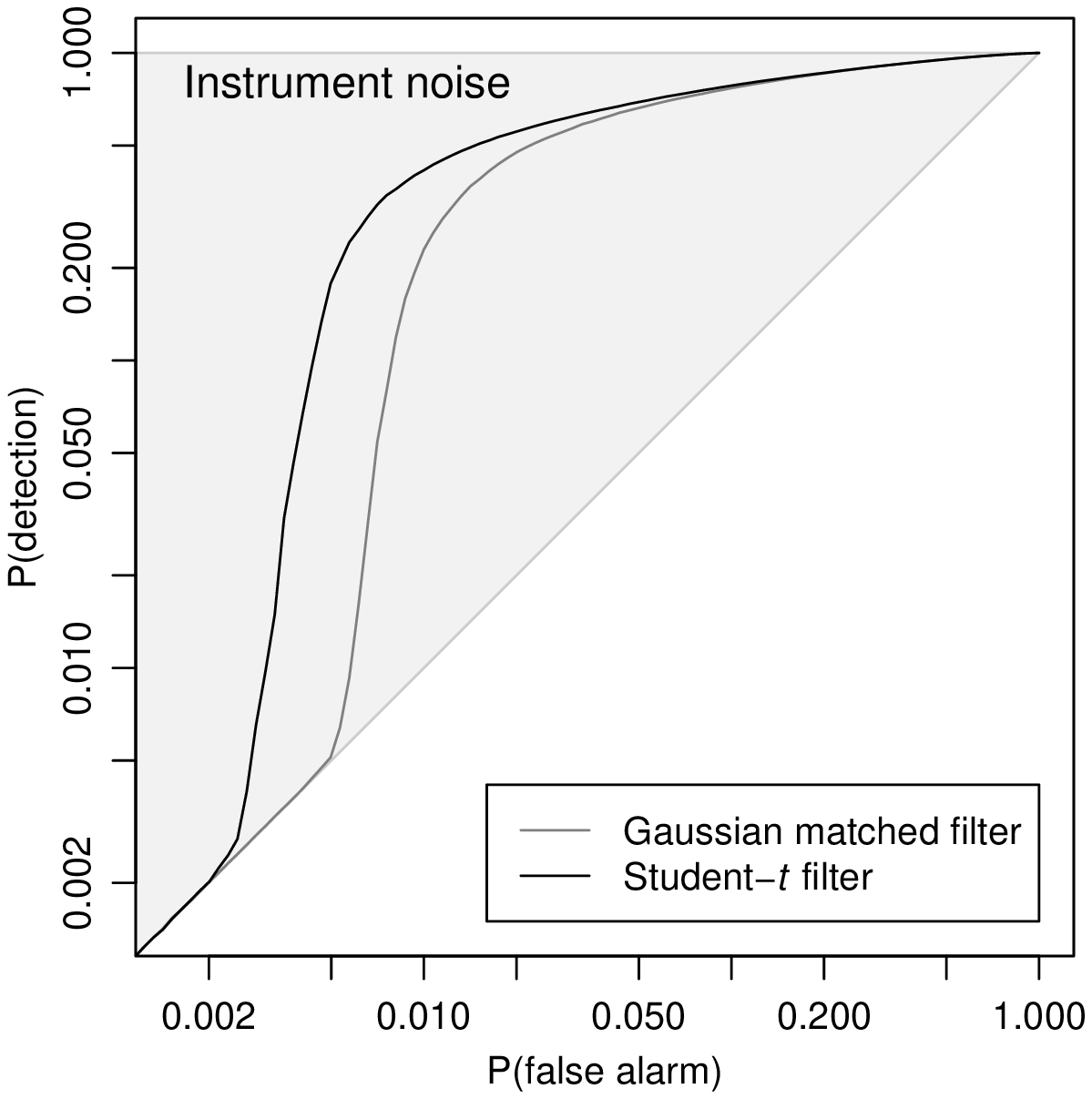}
  \caption{ROC curves for the Gaussian and the Student\mbox{-}$t$
           detection statistics in both data scenarios.  The shaded
           area marks the region where any sensible detection
           statistic (one that is not worse than mere guessing) should
           lie.}
  \label{fig:ROC}
\end{figure*}

\subsection{Filtering setup}
For each piece of data, the likelihood ratio is maximized over phase
and amplitude for given combinations of time and mass parameter
values, where the evaluated time points were
$t_c\in\{6.50,\,6.55,\,\ldots,\,7.50\}$ and the considered masses were
$\eta=0.25$, $m_c \in \{3.0,\,3.1,\,\ldots,\,6.0\}$.  The injected
signal's parameter values always were among the grid points maximized
over, so that signal/template mismatch considerations are not of
concern here.  On the technical side, this is implemented in a loop
over template waveforms (corresponding to different mass parameters)
and time points. At each mass/time combination, computation of the
conditionally maximized Gaussian likelihood ratio amounts to computing
an inner product / quadratic form (see
Sec.~\ref{sec:GaussLikeliMaxi}), while maximizing the conditional
Student\mbox{-}$t$ likelihood requires iterating over several such
least-squares fits within the EM~algorithm (see
Sec.~\ref{sec:StudentLikeliMaxi}). The EM~iterations were terminated
whenever the improvement in logarithmic likelihood over the previous
iteration fell below~$10^{-6}$\@. In this example setting, this lead
to an average number of four EM~iterations for each conditional
likelihood maximization in both noise scenarios.  The eventual
maximized likelihood then is given by the overall maximum over the
conditional maxima, and as the detection statistic we use the
maximized likelihood ratio $\frac{p(d|\hat{\theta})}{p(d|\vec{0})}$.
The algorithm used was essentially the one described in
Appendix~\ref{sec:implementationStudentGeneral}.

\subsection{Simulation results}
Figure~\ref{fig:detectionStats} shows resulting detection statistic
values (maximized likelihood ratios) under the Gaussian and the
Student\mbox{-}$t$ models both when a signal is injected as well as
when he data are noise only. The signal injections here were all done
at the same amplitude relative to the noise spectrum
(SNR~$\varrho=5.257$).  In general, both detection statistics are very
similar; the Student\mbox{-}$t$ likelihood ratio tends to turn out
slightly lower than the Gaussian one, in particular in the case of
real interferometer noise.

The question of to what extent these differences affect the ability to
discriminate signals from noise will be approached by considering the
receiver operating characteristic (ROC) curves.  ROC curves are based
on the detection statistics' (here: empirical) distributions. Placing
different detection thresholds on a detection statistic yields a
corresponding false alarm probability (based on the distribution under
the noise-only hypothesis) as well as a detection probability (based
on the distribution under the particular signal hypothesis). The ROC
curve illustrates these combinations over varying threshold values
\cite{Fawcett2006}.

\begin{figure}
  \centering
  \includegraphics[width=0.75\columnwidth]{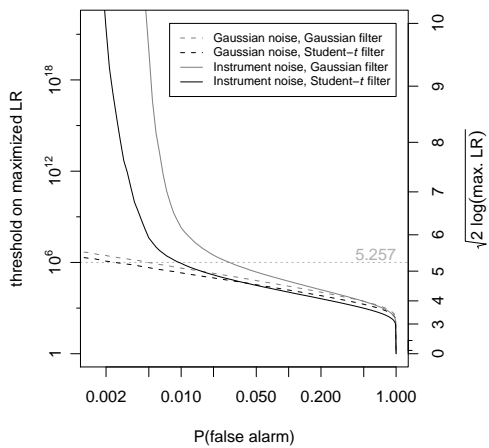}
  \caption{Detection thresholds on the maximized likelihood ratio (the
           detection statistic), corresponding to certain false-alarm
           probabilities. These thresholds are based on the detection
           statistic's distribution in the absence of a signal. The
           horizontal line indicates the injected signals' SNR (see
           also Fig.~\ref{fig:ROC}).}
  \label{fig:detectionThresholds}
\end{figure}

Figure~\ref{fig:ROC} shows ROC curves for the Gaussian and the
Student\mbox{-}$t$ filter for both noise cases.  In the case of
simulated Gaussian noise, both detection statistics perform almost
identically. For real instrument noise on the other hand, the
Student\mbox{-}$t$ model is able to provide a significantly greater
detection probability especially at low false-alarm probabilities.  A
remarkable feature of the ROC curves for instrumental noise is that for
very low false alarm probabilities both filters eventually perform as
poorly as mere guessing. The Student\mbox{-}$t$ filter is able to
sustain its discriminating power for lower false alarm rates, though.
This effect is connected to the frequency of noise outliers
(``glitches'') in the data, leading to very large detection statistic
values even in the absence of a signal.
Figure~\ref{fig:detectionThresholds} shows the corresponding detection
thresholds as a function of false-alarm probabilities.  The point
where the detection threshold reaches the injected signals' SNR is
where the corresponding detection probability is $\approx 50\%$.  One
can see that, due to the heavy-tailed distribution of detection
statistics in the case of actual instrument noise, the detection
threshold necessary for low false-alarm probabilities very quickly
grows beyond values that could obviously be attributed to be due to
the signal injections considered here; the rate of noise transients of
``SNR'' greater than the injections' SNR exceeds the false alarm rate
(in a realistic search, some of these might actually be vetoed
beforehand).
This effect is very obvious here also because signal injections were
done only at a single SNR, but it will of course persist for other SNR
distributions---assuming other SNR distributions for injections will
affect the detection probability, but not the detection threshold,
i.e., the detection procedure itself.

The exact relative performance of both methods of course depends on
the details of the particular detection problem, the kind of signal
searched for, the parameter space, noise characteristics, data
conditioning, and tuning parameters.
The ROC curves shown above are based on a particular, artificial
signal population, but their general features persist in a number of
additional simulations not shown here, for a range of degrees-of-freedom settings,
injection SNRs, data from a different instrument, and data from a
different time period.

\section{Conclusions}
We introduced a generalization of the matched filter that is commonly
applied in signal detection problems.  The Student\mbox{-}$t$ filter
is derived as a maximum-likelihood detection method that is based on a
Student\mbox{-}$t$ distribution for the noise, rather than a Gaussian
distribution, which would again yield the common matched filter
instead. On the technical side, it generalizes a least-squares method
to an adaptive variety.  While a ``Gaussian'' matched filter is
certainly appropriate when the assumption of stationary Gaussian noise
and a known spectrum is met, there are several ways to motivate the
Student\mbox{-}$t$ filter as a robust alternative when these
assumptions are violated:
  (i)~``\textsl{theoretically}'': the Student\mbox{-}$t$ model allows for uncertainty in the PSD, heavier-tailed noise and outliers;
 (ii)~``\textsl{heuristically}'': the resulting adaptive least-squares method is less outlier-sensitive; or
(iii)~``\textsl{pragmatically}'': the filter may turn out more effective in practice, as in the realistic example shown above.
Besides that, being a generalization of the (Gaussian) matched filter,
it should generally be able to perform as well or better.  The
question of course is whether the gain in detection efficiency is
worth the additional implementation, tuning and computational effort.
The difference in computational cost for deriving both detection
statistics suggests that a combined, hierarchical search strategy may
also be worth considering.

In the example shown above, the Student\mbox{-}$t$ model's
degrees-of-freedom parameter was treated as a single constant. In the
context of gravitational-wave interferometric data, this is an
oversimplification; a study of actual instrument noise shows that the
Fourier-domain data's tail behavior clearly depends on the frequency
\cite{Waldman2011,Roever2011-LigoTechReport}. 
Accounting for this effect in an actual search by fitting
individual~$\nu_j$ parameters for different frequency ranges may yield
a significant improvement. It may also make sense to specify the
degrees-of-freedom parameter dependent on additional information, like
e.g.\ the data quality category \cite{SlutskyEtAl2010}.

It will be interesting to study the Student\mbox{-}$t$ filter's
performance in a realistic search for gravitational-wave signals, in
conjunction with the existing infrastructure (data quality flags,
additional vetoes, etc.) and in comparison with the conventional
matched filter \cite{AllenEtAl2005,Brown2005}. We are also
investigating the use of the Student\mbox{-}$t$ model in the context
of Bayesian model selection \cite{VeitchVecchio2008a}.  Here it may
again yield a more robust discriminator for actual signals against
noise; on the computational side this problem is based on integration
of the likelihood, rather than maximization, and we do not expect a
difference in computational cost between Gaussian and
Student\mbox{-}$t$ models.
We expect the Student\mbox{-}$t$ filtering procedure to be also useful
in many other signal-processing contexts, wherever robustness or
uncertainty in the power spectrum is an issue.

\begin{acknowledgments}
  The author wishes to thank Nelson Christensen, Drew Keppel, Karsten
  L\"{u}bke, Renate Meyer, and Reinhard Prix for fruitful discussions
  at various stages of this work, 
  and the LIGO Scientific Collaboration (LSC) 
  for providing the data used here.  
  The author gratefully acknowledges the support of the United States
  National Science Foundation for the construction and operation of
  the LIGO Laboratory and the Science and Technology Facilities
  Council of the United Kingdom, the Max-Planck-Society, and the State
  of Niedersachsen/Germany for support of the construction and
  operation of the GEO600 detector. The author also gratefully
  acknowledges the support of the research by these agencies and by
  the Australian Research Council, the International Science Linkages
  program of the Commonwealth of Australia, the Council of Scientific
  and Industrial Research of India, the Istituto Nazionale di Fisica
  Nucleare of Italy, the Spanish Ministerio de Educaci\'on y Ciencia,
  the Conselleria d'Economia, Hisenda i Innovaci\'o of the Govern de
  les Illes Balears, the Royal Society, the Scottish Funding Council,
  the Scottish Universities Physics Alliance, The National Aeronautics
  and Space Administration, the Carnegie Trust, the Leverhulme Trust,
  the David and Lucile Packard Foundation, the Research Corporation,
  and the Alfred P. Sloan Foundation.
\end{acknowledgments}

\setcounter{equation}{0}
\begin{appendix}
\section*{Appendix}
\setcounter{section}{1}
\subsection{Discrete Fourier transform}\label{sec:DFTdefinition}
The Fourier transform convention used in this paper is specified
below; it is defined for a real-valued function~$h$ of time~$t$,
sampled at $N$ discrete time points, at a sampling rate of
$\frac{1}{\Delta_t}$, and it maps from
\begin{equation} \label{eqn:DFTtimedomain}
  \{h(t)\in\realline:\; t=0,\Delta_t,2\Delta_t,\ldots,(N-1)\Delta_t\}
\end{equation}
to a function of frequency~$f$
\begin{equation} \textstyle
  \{\tilde{h}(f)\in\complexnumb:\; f=0,\Delta_f,2\Delta_f,\ldots,(N-1)\Delta_f\},
\end{equation}
where $\Delta_f = \frac{1}{N\Delta_t}$ and
\begin{equation}\label{eqn:DftDefinition}
  \tilde{h}(f) \;=\; \sum_{j=0}^{N-1} h(j\Delta_t) \,\exp(-2 \pi \imag j \Delta_t f)
\end{equation}
\cite{RoeverMeyerChristensen2011}.

\subsection{Applying the EM algorithm }\label{sec:EMApp}
\subsubsection{Preliminaries}
The \textsl{expectation-maximization (EM) algorithm} is required for
maximizing the Student\mbox{-}$t$ likelihood; see
Sec.~\ref{sec:StudentLikeliMaxi}.  What is desired is the maximum of
the marginal likelihood $p(d|\theta)$, which is equivalent to the
marginal density $p(\theta|d)$ when assuming a uniform prior
distribution on $\theta$.  What is required in order to apply the
EM~algorithm are expressions involving the marginalized
$\sigma_j^2$~parameters, namely, the conditional distribution
$\prob(\vec{\sigma}^2|\theta,d)$ and the joint density
$p(\theta,\vec{\sigma}^2|d)$.  The EM~algorithm will then iteratively
maximize the likelihood function by performing alternating
``expectation'' and ``maximization'' steps
\cite{BDA,DempsterLairdRubin1977}.

The conditional posterior distribution $\prob(\sigma_j^2|\theta,d)$ of
the $j$th variance parameter~$\sigma_j^2$ for given data and
signal~$s_\theta$ is a scaled inverse $\chi^2$~distribution,
\begin{equation}
  \invchisq\biggl(\nu_j+2, \frac{\nu_j S_1(f_j)_j+4\frac{\Delta_t}{N}\bigl|\tilde{d}(f_j)-\tilde{s}_\theta(f_j)\bigr|^2}{\nu_j+2}\biggr)
\end{equation}
\cite{RoeverMeyerChristensen2011} 
with probability density
function
\begin{equation}
  f(\sigma_j^2) 
  \propto
  \bigl(\sigma_j^2\bigr)^{-\frac{\nu_j+4}{2}}
  \exp\biggl(-\frac{\frac{\nu_j}{2}S_1(f_j)+4\frac{\Delta_t}{N}\bigl|\tilde{d}(f_j)\!-\!\tilde{s}_\theta(f_j)\bigr|^2}{2\sigma_j^2}\biggr)
\end{equation}
\cite{RoeverMeyerChristensen2011}.

The conditional distribution of the data~$d$ for given
variances~$\vec{\sigma}^2$ and signal parameters~$\theta$,
$\prob(y|\theta,\vec{\sigma}^2)$, is Gaussian
\cite{RoeverMeyerChristensen2011},
and the variance parameters' prior, $\prob(\vec{\sigma}^2)$, 
again was $\invchisq$ \cite{RoeverMeyerChristensen2011}.
The joint conditional density of $\theta$ and $\vec{\sigma}^2$ 
for given data~$d$ is given by
\begin{eqnarray}
  &&\log\bigl(p(\theta, \sigma^2 | y) \bigr)
  \;\propto\;
  \log\bigl(p(y|\theta,\sigma^2) \times p(\theta,\sigma^2)\bigr)\\
  &\propto&
  -\sum_j \Bigl(\log(\sigma_j^2) + \textstyle \frac{4\frac{\Delta_t}{N}\bigl|\tilde{d}(f_j)-\tilde{s}_\theta(f_j)\bigr|^2}{2\sigma_j^2} \Bigr)
  \nonumber \\
  &&
  -\sum_j \Bigl((1+\textstyle\frac{\nu_j}{2})\log(\sigma_j^2) + \textstyle\frac{\nu_jS_1(f_j)}{2\sigma_j^2}\Bigr)\\
  &=&
  -\sum_j \Bigl((2+\textstyle\frac{\nu_j}{2})\log(\sigma_j^2) + \textstyle\frac{\nu_jS_1(f_j) + 4\frac{\Delta_t}{N}\bigl|\tilde{d}(f_j)-\tilde{s}_\theta(f_j)\bigr|^2}{2\sigma_j^2} \Bigr)\label{eqn:jointPosterior}
\end{eqnarray}
\cite{RoeverMeyerChristensen2011}.

\subsubsection{The E step}
For the EM~algorithm's \textsl{expectation} step, one needs to evaluate the
conditional posterior expectation
\begin{eqnarray}
  &&\expect_{\prob(\sigma^2|\theta=\theta_0,y)}\bigl[ \log\bigl(p(\theta, \sigma^2 | y) \bigr) \bigr]\nonumber\\
  &=& 
  \int  \log\bigl(p(\theta, \sigma^2 | y) \bigr) \, p(\sigma^2|\theta\!=\!\theta_0,y) \, \differential\sigma^2
\end{eqnarray}
as a function of $\theta$ for some given $\theta_0$
\cite{BDA}.
Here,
\begin{widetext}
\begin{eqnarray}
  && \int  \log\bigl(p(\theta, \sigma^2 | y) \bigr) \,p(\sigma^2|\theta\!=\!\theta_0,y) \,\differential\sigma^2\\
  &\propto& -\sum_j 
    \int\Bigl((2+\textstyle\frac{\nu_j}{2})\log(\sigma^2) 
                 + \textstyle\frac{\nu_jS_1(f_j) + 4\frac{\Delta_t}{N}\bigl|\tilde{d}(f_j)-\tilde{s}_\theta(f_j)\bigr|^2}{2\sigma_j^2} \Bigr)
  \nonumber \\
  && \qquad \qquad
  \times
           \Bigl( \bigl(\sigma^2\bigr)^{-(2+\frac{\nu_j}{2})}
                  \exp\Bigl(\textstyle\frac{\nu_jS_1(f_j)+4\frac{\Delta_t}{N}\bigl|\tilde{d}(f_j)-\tilde{s}_{\theta_0}(f_j)\bigr|^2}{2\sigma^2}\Bigr)\Bigr)
    \differential \sigma_j^2\\
  &\propto& -\sum_j 
    {\textstyle\frac{4\frac{\Delta_t}{N}\bigl|\tilde{d}(f_j)-\tilde{s}_\theta(f_j)\bigr|^2}{2}}
    \times\int\textstyle\frac{1}{\sigma_j^2}
           \Bigl( \bigl(\sigma^2\bigr)^{-(2+\frac{\nu_j}{2})}
                  \exp\Bigl(\textstyle\frac{\nu_jS_1(f_j)+4\frac{\Delta_t}{N}\bigl|\tilde{d}(f_j)-\tilde{s}_{\theta_0}(f_j)\bigr|^2}{2\sigma^2}\Bigr)\Bigr)
    \differential \sigma_j^2,
\end{eqnarray}
\begin{eqnarray}
  \mbox{where}\quad\int\textstyle\frac{1}{\sigma_j^2}
             \overbrace{\Bigl( \bigl(\sigma^2\bigr)^{-(2+\frac{\nu_j}{2})}
                    \exp\Bigl(\textstyle\frac{\nu_jS_1(f_j)+4\frac{\Delta_t}{N}\bigl|\tilde{d}(f_j)-\tilde{s}_{\theta_0}(f_j)\bigr|^2}{2\sigma^2}\Bigr)\Bigr)}^{(\ast)}
    \differential \sigma_j^2
  &=& 
  \frac{\nu_j+2}{\nu_jS_1(f_j)+4\frac{\Delta_t}{N}\bigl|\tilde{d}(f_j)-\tilde{s}_{\theta_0}(f_j)\bigr|^2},
\end{eqnarray}
\end{widetext}
since the term marked by the asterisk~$(\ast)$ is the density function
of an
$\invchisq\bigl(\nu_j+2,\frac{\nu_jS_1(f_j)+4\frac{\Delta_t}{N}\bigl|\tilde{d}(f_j)-\tilde{s}_{\theta_0}(f_j)\bigr|^2}{\nu_j+2}\bigr)$
probability distribution, so that
\begin{eqnarray}
  &&
  \int  \log\bigl(p(\beta, \sigma^2 | y) \bigr) \,p(\sigma^2|\beta\!=\!\beta_0,y) \,\differential\sigma^2 \nonumber
  \\
  &\propto&
  -{\textstyle\frac{1}{2}}\sum_j {\textstyle \frac{4\frac{\Delta_t}{N}\bigl|\tilde{d}(f_j)-\tilde{s}_\theta(f_j)\bigr|^2}{\frac{\nu_j}{\nu_j+2}S_1(f_j)+\frac{1}{\nu_j+2}\bigl(4\frac{\Delta_t}{N}\bigl|\tilde{d}(f_j)-\tilde{s}_{\theta_0}(f_j)\bigr|^2\bigr)}}
  \\
  &=&
  -{\textstyle\frac{1}{2}}\sum_j {\textstyle \frac{\bigl|\tilde{d}(f_j)-\tilde{s}_\theta(f_j)\bigr|^2}{\frac{N}{4\Delta_t} \Bigl(\frac{\nu_j}{\nu_j+2}\,S_1(f_j)\,+\,\frac{2}{\nu_j+2}\frac{2\Delta_t}{N}\bigl|\tilde{d}(f_j)-\tilde{s}_{\theta_0}(f_j)\bigr|^2\Bigr)}}\label{eqn:EM_expectation1}
  \\
  &=:& 
  \mathcal{E}(\theta_0, \theta).\nonumber
\end{eqnarray}

\subsubsection{The M step}
In the EM~algorithm's \textsl{maximization} step, the above
expectation~$\mathcal{E}(\theta_0, \theta)$
(\ref{eqn:EM_expectation1}) needs to be maximized with respect to the
parameter~$\theta$.  The parameter value maximizing the expectation
then constitutes the next iteration's ``new'' $\theta_0$ value, for
which then the expectation again is maximized, and so forth
\cite{BDA}.  As one can see from expression
(\ref{eqn:EM_expectation1}), maximization of the expectation again
amounts to minimizing weighted least-squares, as in the Gaussian
matched filter described above.

\subsection{Pseudocode matched and Student-t filters}
\label{sec:PseudocodeApp}
\subsubsection{Preliminaries}
This section sketches actual implementations of Student\mbox{-}$t$ and
(Gaussian) matched filters in comparison.  In the following, we will
use essentially the same conventions as before; we will be considering
a time series~$d$ of length~$N$, sampled at a sampling interval
of~$\Delta_t$. The signal waveform here is assumed to be a linear
combination of a sine and a cosine component
($s_{\mathrm{s},\theta}$, $s_{\mathrm{c},\theta}$), it has an
associated arrival time parameter, and possibly additional
parameters~$\theta$ (as in (\ref{eqn:AllenEtAlFilter})).  Additional
waveform parameters (other than amplitude, phase, and time) are then
commonly treated by running several matched filters corresponding to
different values of~$\theta$.  The generalization to the case of more
than two linear signal components should be straightforward.  The
profile likelihood will be evaluated along a discrete grid of time
points $\tau_i$ ($i=1,\ldots,m$), where the special case of
$\tau_i=i\Delta_t$ and $m=N$ is of particular interest.  The filter's
output each time is a single number, the maximized (logarithmic)
likelihood ratio of signal vs.\ no-signal models. We will be making
use of the inner product / quadratic form notation $\langle a, b; S
\rangle$ as defined in~(\ref{eqn:innerProductSum}).
Implementations of the algorithms sketched in
Sec.~\ref{sec:implementationGaussianEfficient} and
\ref{sec:implementationStudentEfficient} are also provided in
\cite{bspec2011}.

\subsubsection{The ``Gaussian'' matched filter: general implementation}
\label{sec:implementationGaussianGeneral}
The first algorithm (Table~\ref{alg:Gauss1}) is a ``naive''
matched-filter implementation that maximizes the likelihood
(\mbox{-}ratio) over a given grid of $m$~time points ($\tau$).  The
algorithm mainly consists of a loop over time points, where for each
time point the (conditional) likelihood is maximized over amplitude
and phase.  In order to match signal and data for a certain signal
arrival time, the data~$d$ are time shifted against the signal
waveforms~$s_{\mathrm{s}/\mathrm{c}}$.  The eventual result is the
profile likelihood evaluated at the specified time points, the maximum
of which then constitutes the generalized likelihood ratio detection
statistic that is returned.
\begin{table}
\caption{Matched filter, general implementation.}
\label{alg:Gauss1}
\begin{ruledtabular}
\begin{tabular}{c}
\begin{minipage}{\columnwidth}
\begin{algorithmic}[5]
  \STATE $\mathtt{normS} \algAssign \langle s_{\mathrm{s},\theta} ,\;  s_{\mathrm{s},\theta} ;\; S_1\rangle$
  \STATE $\mathtt{normC} \algAssign \langle s_{\mathrm{c},\theta} ,\;  s_{\mathrm{c},\theta} ;\; S_1\rangle$
  \FOR[loop over time points:]{$(i=1,\ldots,m)$} 
  \FOR[time shift the data:]{$(j=0,\ldots,N/2)$} 
  \STATE $\tilde{d}_j^\prime \algAssign \tilde{d}_j \times \exp(2\pi\mathrm{i}f_j\tau_i)$
  \ENDFOR
  \STATE $\mathtt{prodS} \algAssign \langle s_{\mathrm{s},\theta} ,\; d^\prime ;\; S_1\rangle$
  \STATE $\mathtt{prodC} \algAssign \langle s_{\mathrm{c},\theta} ,\; d^\prime ;\; S_1\rangle$
  \STATE \COMMENT{compute log-likelihood ratio / profile likelihood:}
  \STATE $\mathtt{maxLLR}[i] \algAssign (\mathtt{prodS})^2/\mathtt{normS} + (\mathtt{prodC})^2/\mathtt{normC}$
  \ENDFOR
  \STATE \textbf{return} $\max(\mathtt{maxLLR})$
\end{algorithmic}
\end{minipage}
\end{tabular}
\end{ruledtabular}
\end{table}

\begin{table}[b]
\caption{Matched filter, efficient implementation.}
\label{alg:Gauss2}
\begin{ruledtabular}
\begin{tabular}{c}
\begin{minipage}{\columnwidth}
\begin{algorithmic}[5]
  \STATE $\mathtt{normS} \algAssign \langle s_{\mathrm{s},\theta} ,\;  s_{\mathrm{s},\theta} ;\; S_1\rangle$
  \STATE $\mathtt{normC} \algAssign \langle s_{\mathrm{c},\theta} ,\;  s_{\mathrm{c},\theta} ;\; S_1\rangle$
  \FOR[correlate data and signals:]{$(j=0,\ldots,(N-1))$} 
  \STATE $\mathtt{corS}[j+1] \algAssign \tilde{d}_j \times \tilde{s}_{\mathrm{s},\theta,j}^\ast \,/\, S_1(f_j)$
  \STATE $\mathtt{corC}[j+1] \algAssign \tilde{d}_j \times \tilde{s}_{\mathrm{c},\theta,j}^\ast \,/\, S_1(f_j)$
  \ENDFOR
  \STATE \COMMENT{apply Fourier transforms:}
  \STATE $\mathtt{FTS} \algAssign \mathrm{DFT}(\mathtt{corS})$
  \STATE $\mathtt{FTC} \algAssign \mathrm{DFT}(\mathtt{corC})$
  \FOR[profile likelihood (\mbox{-}ratio):]{$(i=1,\ldots,N)$} 
  \STATE $\mathtt{maxLLR}[i] \algAssign \bigl(\frac{\Delta_t}{N}\bigr)^2\, \Bigl(\frac{(\mathtt{FTS}[N+1-i])^2}{\mathtt{normS}} + \frac{(\mathtt{FTC}[N+1-i])^2}{\mathtt{normC}}\Bigr)$ \label{algline:Gauss2LLR}
  \ENDFOR
  \STATE \textbf{return} $\max(\mathtt{maxLLR})$ \label{algline:Gauss2Maximization}
\end{algorithmic}
\end{minipage}
\end{tabular}
\end{ruledtabular}
\end{table}

\subsubsection{The ``Gaussian'' matched filter: efficient implementation}
\label{sec:implementationGaussianEfficient}
If the time points to be maximized over are taken to be the same as
the data time series' points ($\tau_i=i\Delta_t,\,i=1,\ldots,m=N$),
then the matched-filtering procedure may be implemented much more
efficiently.  The algorithm shown in Table~\ref{alg:Gauss2} will give
identical results to the previous, but it is more efficient as it
takes advantage of a Fourier transform to essentially maximize over
amplitude, phase, and time simultaneously (see also
Sec.~\ref{sec:commonImplementation}).  In practice, one may want to
restrict the profile likelihood maximization
(line~\ref{algline:Gauss2Maximization}) to the subset of sensible time
shifts that do not ``wrap'' the signal circularly around the data's
end points.  Instead of a Fourier transform, one could also implement
an inverse Fourier transform and would then also not need to
time-reverse the result's indices (line~\ref{algline:Gauss2LLR}).

\subsubsection{The Student-t filter: general implementation}
\label{sec:implementationStudentGeneral}
This algorithm (see Table~\ref{alg:Student1}) again is a ``general''
version of the Student\mbox{-}$t$ filter, analogous to the general
matched filter (Sec.~\ref{sec:implementationGaussianGeneral}), where
the set of time points~$\tau$ is not restricted. The EM~algorithm here
is applied at the level of each single amplitude/phase maximization
conditional on some time shift~$\tau_i$.  
\begin{table}[t]
\caption{Student\mbox{-}$t$ filter, general implementation.}
\label{alg:Student1}
\begin{ruledtabular}
\begin{tabular}{c}
\begin{minipage}{\columnwidth}
\begin{algorithmic}[5]
\STATE $\mathtt{LL0} \algAssign \log(p(d, S_1, \nu))$ \COMMENT{log-likelihood noise-only model}
\FOR[loop over time points:]{$(i=1,\ldots,m)$} 
\FOR[time shift the data:]{$(j=0,\ldots,N/2)$} 
\STATE $\tilde{d}_j^\prime$ $=$ $\tilde{d}_j\times \exp(2\pi\mathrm{i}f_j\tau_i)$
\ENDFOR
\STATE \COMMENT{EM-iterations:}
\STATE $k \algAssign 1$; $\quad\Delta_{\mathtt{LLR}} \algAssign 1$; $\quad\mathtt{LLRprev} \algAssign 0$; $\quad S_1^\star \algAssign S_1$
\WHILE{$(\Delta_{\mathtt{LLR}} > \Delta_{\mathrm{max}})$ \textbf{and} $(k\leq k_{\mathrm{max}})$}
\STATE $\mathtt{normS} \algAssign \langle s_{\mathrm{s},\theta} ,\;  s_{\mathrm{s},\,\theta} ;\; S_1^\star\rangle$
\STATE $\mathtt{normC} \algAssign \langle s_{\mathrm{c},\theta} ,\;  s_{\mathrm{c},\,\theta} ;\; S_1^\star\rangle$
\STATE $\texttt{prodS} \algAssign \langle s_{\mathrm{s},\theta} ,\; d^\prime ;\; S_1^\star\rangle$
\STATE $\texttt{prodC} \algAssign \langle s_{\mathrm{c},\theta} ,\; d^\prime ;\; S_1^\star\rangle$
\STATE $\hat{\beta}_{\mathrm{s}} \algAssign \mathtt{prodS} / \mathtt{normS}$
\STATE $\hat{\beta}_{\mathrm{c}} \algAssign \mathtt{prodC} / \mathtt{normC}$
\STATE $\hat{n} \algAssign d^\prime - \bigl(\hat{\beta}_{\mathrm{s}} s_{\mathrm{s},\theta} + \hat{\beta}_{\mathrm{c}} s_{\mathrm{c},\theta}\bigr)$ \COMMENT{vector of noise residuals}
\STATE $\mathtt{LL1} \algAssign \log(p(\hat{n}, S_1, \nu))$ \COMMENT{log-likelihood signal model}
\STATE $\texttt{LLR} \algAssign \texttt{LL1} - \mathtt{LL0}$ \COMMENT{log-likelihood ratio}
\STATE $\Delta_{\mathtt{LLR}} \algAssign \mathtt{LLR} - \mathtt{LLRprev}$
\STATE $\mathtt{LLRprev} \algAssign \mathtt{LLR}$
\FOR[adapt the spectrum:]{($j=0,\ldots,N/2$)}
\STATE $S_1^\star(f_j) \algAssign \frac{\nu_j}{\nu_j+2}\,S_1(f_j)
                               + \frac{2}{\nu_j+2} \frac{2\Delta_t}{N}\, \bigl|\hat{\tilde{n}}_j\bigr|^2$
\ENDFOR
\STATE $k \algAssign k+1$
\ENDWHILE
\STATE $\mathtt{maxLLR}[i] \algAssign \mathtt{LLR}$ \COMMENT{profile likelihood (\mbox{-}ratio)}
\ENDFOR
\STATE \textbf{return} $\max(\mathtt{maxLLR})$
\end{algorithmic}
\end{minipage}
\end{tabular}
\end{ruledtabular}
\end{table}
The EM component requires
the specification of a threshold~$\Delta_{\mathrm{max}}$ on the
improvement in logarithmic maximized likelihood ratio (e.g.\
$10^{-6}$), and a threshold~$k_{\mathrm{max}}$ on the number of EM
iterations (e.g.\ $100$).  The Student-$t$ likelihood function
\[
p(x, S_1, \nu)
\;\propto\;
\exp\biggl(- \sum_{j} {\textstyle \frac{\nu_j + 2}{2}} \log\biggl[1 + \frac{1}{\nu_j}\,\frac{\bigl|\tilde{x}_j\bigr|^2}{\frac{N}{4\Delta_t}\,S_1(f_j)}\biggr]\biggr)
\] 
(see also (\ref{eqn:StudentTLikeli2})) only needs to be computed up to
a proportionality constant here, as only the likelihood \textsl{ratio}
is of eventual interest.

\begin{table}[t]
\caption{Student\mbox{-}$t$ filter, efficient implementation.}
\label{alg:Student2}
\begin{ruledtabular}
\begin{tabular}{c}
\begin{minipage}{\columnwidth}
\begin{algorithmic}[5]
\STATE $\mathtt{LL0} \algAssign \log(p(d, S_1, \nu))$ \COMMENT{log-likelihood noise-only model}
\STATE \COMMENT{EM-iterations:}
\STATE $k \algAssign 1$; $\quad\Delta_{\mathtt{LLR}} \algAssign 1$; $\quad \mathtt{LLRprev} \algAssign 0$; $\quad S_1^\star \algAssign S_1$
\WHILE{$(\Delta_{\mathtt{LLR}} > \Delta_{\mathrm{max}})$ \textbf{and} $(k\leq k_{\mathrm{max}})$}
\STATE\COMMENT{the ``plain'' matched filter:}
\STATE $\mathtt{normS} \algAssign \langle s_{\mathrm{s},\theta} ,\;  s_{\mathrm{s},\theta} ;\; S_1^\star\rangle$ \label{algline:Student2FilterStart}
\STATE $\mathtt{normC} \algAssign \langle s_{\mathrm{c},\theta} ,\;  s_{\mathrm{c},\theta} ;\; S_1^\star\rangle$
\FOR{$(j=0,\ldots,(N-1))$} 
\STATE $\mathtt{corS}[j+1] \algAssign \tilde{d}_j \times \tilde{s}_{\mathrm{s},\theta,j}^\ast \,/\, S_1^\star(f_j)$
\STATE $\mathtt{corC}[j+1] \algAssign \tilde{d}_j \times \tilde{s}_{\mathrm{c},\theta,j}^\ast \,/\, S_1^\star(f_j)$
\ENDFOR
\STATE $\mathtt{FTS} \algAssign \mathrm{DFT}(\mathtt{corS})$
\STATE $\mathtt{FTC} \algAssign \mathrm{DFT}(\mathtt{corC})$
\FOR{$(i=1,\ldots,N)$} 
\STATE $\mathtt{maxLLR}[i] \algAssign \bigl(\frac{\Delta_t}{N}\bigr)^2\, \Bigl(\frac{(\mathtt{FTS}[N+1-i])^2}{\mathtt{normS}} + \frac{(\mathtt{FTC}[N+1-i])^2}{\mathtt{normC}}\Bigr)$
\ENDFOR \label{algline:Student2FilterEnd}
\STATE\COMMENT{end of ``plain'' matched filter.}
\STATE \COMMENT{Determine best-fitting template, residuals, etc.:}
\STATE $i_{\mathrm{max}} \algAssign \argmax_i \, \mathtt{maxLLR}[i]$
\FOR[time shift the data:]{$(j=0,\ldots,N/2)$} 
\STATE $\tilde{d}_j^\prime \algAssign \tilde{d}_j\times \exp(2\pi\mathrm{i}f_j\tau_{i_{\mathrm{max}}})$
\ENDFOR
\STATE $\mathtt{prodS} \algAssign \langle s_{\mathrm{s},\theta} ,\; d^\prime ;\; S_1^\star\rangle$
\STATE $\mathtt{prodC} \algAssign \langle s_{\mathrm{c},\theta} ,\; d^\prime ;\; S_1^\star\rangle$
\STATE $\hat{\beta}_{\mathrm{s}} \algAssign \mathtt{prodS} / \mathtt{normS}$
\STATE $\hat{\beta}_{\mathrm{c}} \algAssign \mathtt{prodC} / \mathtt{normC}$
\STATE $\hat{n} \algAssign d^\prime - \bigl(\hat{\beta}_{\mathrm{s}} s_{\mathrm{s},\theta} + \hat{\beta}_{\mathrm{c}} s_{\mathrm{c},\theta}\bigr)$ \COMMENT{vector of noise residuals}
\STATE $\mathtt{LL1} \algAssign \log(p(\hat{n}, S_1, \nu))$ \COMMENT{log-likelihood signal model}
\STATE $\mathtt{LLR} \algAssign \mathtt{LL1} - \mathtt{LL0}$ \COMMENT{log-likelihood ratio}
\STATE $\Delta_{\mathtt{LLR}} \algAssign \mathtt{LLR} - \mathtt{LLRprev}$
\STATE $\mathtt{LLRprev} \algAssign \texttt{LLR}$
\FOR[adapt the spectrum:]{$(j=0,\ldots,N/2)$}
\STATE $S_1^\star(f_j) \algAssign \frac{\nu_j}{\nu_j+2}\,S_1(f_j)
                               + \frac{2}{\nu_j+2} \frac{2\Delta_t}{N}\, \bigl|\hat{\tilde{n}}_j\bigr|^2$
\ENDFOR
\STATE $k \algAssign k+1$
\ENDWHILE
\STATE \textbf{return} $\mathtt{LLR}\phantom{)}$
\end{algorithmic}
\end{minipage}
\end{tabular}
\end{ruledtabular}
\end{table}

\subsubsection{The Student-t filter: efficient implementation}
\label{sec:implementationStudentEfficient}
The Student\mbox{-}$t$ filter also may be implemented more efficiently
in case the signal arrival times to maximize over are taken to be the
time points of the original time series
($\tau_i=i\Delta_t,\,i=1,\ldots,m=N$, as in
Sec.~\ref{sec:implementationGaussianEfficient}).  This implementation
(Table~\ref{alg:Student2}) then requires one to move the level at
which the EM-algorithm is applied from the conditional maximization
over amplitude and phase to the joint amplitude/phase/time
maximization; effectively this implementation iteratively runs several
matched filters (see lines
\ref{algline:Student2FilterStart}--\ref{algline:Student2FilterEnd})
while adapting the noise spectrum in between.  It is unclear whether
or how the level at which the EM-algorithm is applied affects the
results; as noted in Sec.~\ref{sec:StudentFilterImplementation}, the
likelihood may be multimodal and different implementations might end
up with differing maximization results, but whether this actually
poses a problem in practice is not obvious. Computationally, this
latter implementation should be much easier, though.
Another difference to note is that while the matched filter allows us
to return the profile likelihood as a function of time (the \emph{SNR
time series}), only the Student\mbox{-}$t$ filter implementation from
Sec.~\ref{sec:implementationStudentGeneral} is able to provide this,
while the more efficient implementation will only return the overall
maximum.

\subsection{The Student-Rayleigh distribution}\label{sec:StudentRayleighApp}
\subsubsection{Relation to the F distribution}
The noise power's probability distribution under the
Student\mbox{-}$t$ model (see~(\ref{eqn:empiricalNoisePower}),
Sec.~\ref{sec:DFSetting}) may be related to Snedecor's
$F$~distribution.  
\begin{figure}[t]
  \includegraphics[width=0.75\columnwidth]{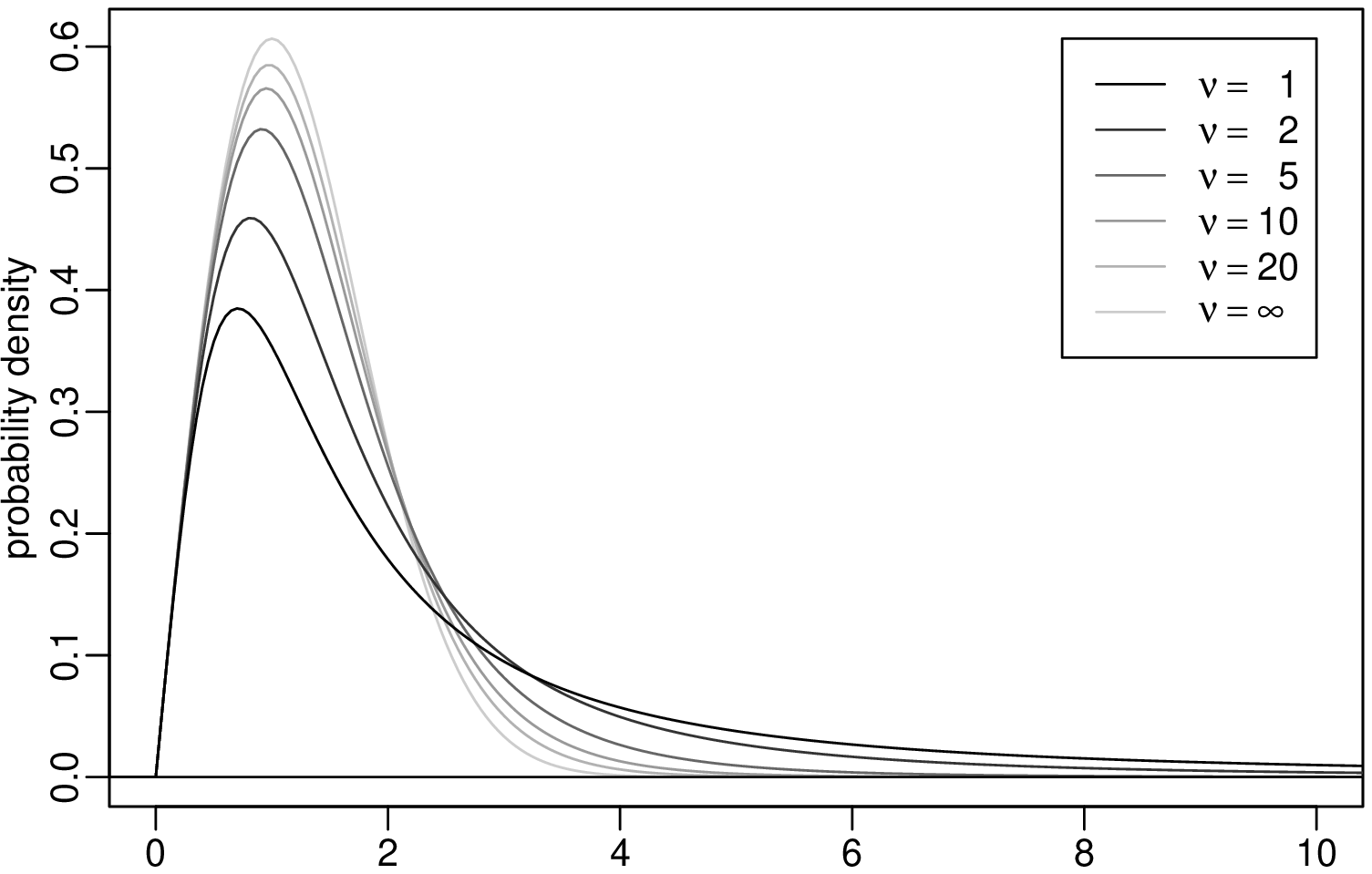}
  \vspace{1ex}
  \includegraphics[width=0.75\columnwidth]{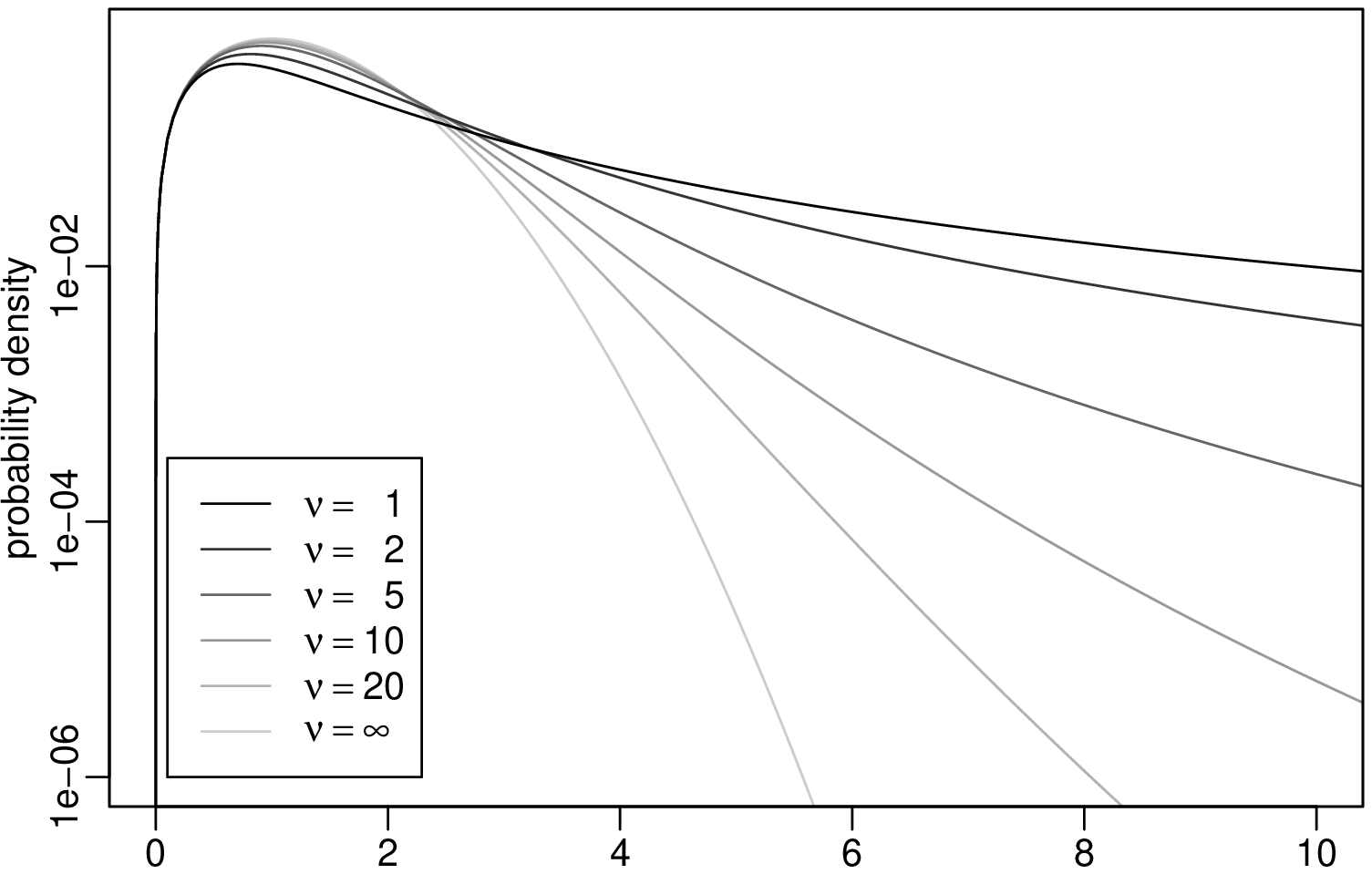}
  \caption{Probability density functions of Student-Rayleigh
           distributions for varying degrees of freedom~$\nu$ and
           fixed scale~$\sigma^2=1$. For $\nu=\infty$, the
           distribution corresponds to the usual (``Gaussian'')
           Rayleigh distribution.}
  \label{fig:SRDensities}
\end{figure}
First, real and imaginary parts of the $j$th element of the discretely
Fourier-transformed vector~$n$ follow a multivariate
(\textsl{bi}variate) Student\mbox{-}$t$ distribution (see
Sec.~\ref{sec:StudentTModel}).  Let $A$ and $B$ be independent
Gaussian random variables with zero mean and standard
deviation~$\sigma$. Furthermore, let $C$ be a $\chi^2_\nu$~distributed
random variable with $\nu$~degrees of freedom. Then the random vector
\[
  \left( \begin{array}{c}
           X \\ Y
         \end{array} \right)
  \;=\;
  \frac{1}{\sqrt{C / \nu}}
  \left( \begin{array}{c}
           A \\ 
           B
         \end{array} \right)
\]
follows a bivariate Student\mbox{-}$t$ distribution with a diagonal
covariance matrix, exactly like the real and imaginary components of
$\tilde{n}(f_j)$ \cite{BDA}.  The root-mean-square figure
corresponding to the power then may be written as
\begin{equation}
  \sqrt{X^2 + Y^2} 
  \;=\;
  \sqrt{2\sigma^2\,\frac{\Bigl(\bigl(\frac{A}{\sigma}\bigr)^2+\bigl(\frac{B}{\sigma}\bigr)^2\Bigr) \, / \, 2}{C \, / \, \nu}}
  \;=\;
  \sqrt{2\sigma^2\, D},
\end{equation}
where the random variable~$D$, being a ratio of
$\chi^2$~distributed random variables that are normalized by
their respective degrees-of-freedom, follows an
$F(2,\nu)$~distribution with $2$ and $\nu$ degrees of freedom
\cite{MGB}.

\subsubsection{Probability density function, etc.}
In the Gaussian noise model (see Sec.~\ref{sec:GaussModel}), the noise
power at the $j$th frequency bin, $\bigl|\tilde{n}(f_j)\bigr|$,
follows a Rayleigh distribution with probability density function
\begin{equation}
  f_{\mathrm{R}}(x|\sigma) \;=\; {\textstyle \frac{x}{\sigma^2}}\, \exp\bigl(\textstyle -\frac{x^2}{2\sigma^2}\bigr),
\end{equation}
where the scale parameter~$\sigma$ is given as
$\sigma = \sqrt{\frac{N}{4\Delta_t}S_1(f_j)}$.

The analogue Student-Rayleigh probability distribution in the
Student\mbox{-}$t$ noise model (see Sec.~\ref{sec:StudentTModel}) is
defined through its density function
\begin{equation}
  f_{\mathrm{SR}}(x|\sigma,\nu) 
  \;=\; 
  {\textstyle \frac{x}{\sigma^2}}\, f_{F(2,\nu)}\bigl(\textstyle \frac{x^2}{2\sigma^2}\bigr),
\end{equation}
where $f_{F(2,\nu)}(\cdot)$ is the probability density function of an
$F(2,\nu)$~distribution with $2$ and $\nu$ degrees of freedom.
Similarly, the cumulative distribution function and quantile function
are given by
\begin{eqnarray}
  F_{\mathrm{SR}}(x|\sigma,\nu) 
  &=&
  F_{F(2,\nu)}\bigl(\textstyle \frac{x^2}{2\sigma^2}\bigr)
  \quad\mbox{and}
  \\
  Q_{\mathrm{SR}}(p|\sigma,\nu) 
  &=&
  \sqrt{2\sigma^2 \, Q_{F(2,\nu)}(p)},
\end{eqnarray}
where $F_{F(2,\nu)}(\cdot)$ and $Q_{F(2,\nu)}(\cdot)$ are the
$F$~distribution's cumulative distribution function and
quantile function.

Figure~\ref{fig:SRDensities} illustrates probability density functions
of Student-Rayleigh probability distributions for varying degrees of
freedom~$\nu$.  For $\nu=\infty$, the distribution corresponds to the
usual (``Gaussian'') Rayleigh distribution. Note, in particular, the
differing tail behavior (analogous to Fig.~\ref{fig:densities}) that
is apparent especially in the logarithmic plot.
\end{appendix}

\bibliography{../../literature/literature}

\begin{thebibliography}{10}

\bibitem{Thorne1987}
K.~S. Thorne.
\newblock Gravitational radiation.
\newblock In S.~W. Hawking and W.~Israel, editors, {\em 300~years of
  gravitation}, chapter~9, pages 330--358. Cambridge University Press,
  Cambridge, 1987.

\bibitem{Schutz1999}
B.~F. Schutz.
\newblock Gravitational wave astronomy.
\newblock {\em Classical and Quantum Gravity}, 16(12A):A131--A156, December
  1999.

\bibitem{RoeverMeyerChristensen2011}
C.~R\"{o}ver, R.~Meyer, and N.~Christensen.
\newblock Modelling coloured residual noise in gravitational-wave signal
  processing.
\newblock {\em Classical and Quantum Gravity}, 28(1):015010, January 2011.

\bibitem{Turin1960}
G.~L. Turin.
\newblock An introduction to matched filters.
\newblock {\em IRE Transactions on Information Theory}, 6(3):311--329, June
  1960.

\bibitem{ChoudhouriGhosalRoy2004a}
N.~Choudhuri, S.~Ghosal, and A.~Roy.
\newblock Contiguity of the {W}hittle measure for a {G}aussian time series.
\newblock {\em Biometrika}, 91(4):211--218, 2004.

\bibitem{Finn1992}
L.~S. Finn.
\newblock Detection, measurement, and gravitational radiation.
\newblock {\em Physical Review~{D}}, 46(12):5236--5249, December 1992.

\bibitem{MGB}
A.~M. Mood, F.~A. Graybill, and D.~C. Boes.
\newblock {\em Introduction to the theory of statistics}.
\newblock McGraw-Hill, New York, 3rd edition, 1974.

\bibitem{BDA}
A.~Gelman, J.~B. Carlin, H.~Stern, and D.~B. Rubin.
\newblock {\em Bayesian data analysis}.
\newblock Chapman \& Hall / CRC, Boca Raton, 1997.

\bibitem{Berger}
J.~O. Berger.
\newblock {\em Statistical decision theory and {B}ayesian analysis}.
\newblock Springer-Verlag, 2nd edition, 1985.

\bibitem{PrixKrishnan2009}
R.~Prix and B.~Krishnan.
\newblock Targeted search for continuous gravitational waves: {B}ayesian versus
  maximum-likelihood statistics.
\newblock {\em Classical and Quantum Gravity}, 26(20):204013, October 2009.

\bibitem{Searle2008}
A.~C. Searle.
\newblock {M}onte-{C}arlo and {B}ayesian techniques in gravitational wave burst
  data analysis.
\newblock {\em Arxiv preprint 0804.1161 [gr-qc]}, April 2008.

\bibitem{RoeverEtAl2009}
C.~R\"{o}ver, M.-A. Bizouard, N.~Christensen, H.~Dimmelmeier, I.~S. Heng, and
  R.~Meyer.
\newblock Bayesian reconstruction of gravitational wave burst signals from
  simulations of rotating stellar core collapse and bounce.
\newblock {\em Physical Review~{D}}, 80(10):102004, November 2009.

\bibitem{CannonEtAl2010}
K.~Cannon, A.~Chapman, C.~Hanna, D.~Keppel, A.~C. Searle, and A.~Weinstein.
\newblock Singular value decomposition applied to compact binary coalescence
  gravitational-wave signals.
\newblock {\em Physical Review~{D}}, 82(4):044025, August 2010.

\bibitem{JaranowskiKrolakSchutz1998}
P.~Jaranowski, A.~Kr\'{o}lak, and B.~Schutz.
\newblock Data analysis of gravitational-wave signals from spinning neutron
  stars: {T}he signal and its detection.
\newblock {\em Physical Review~{D}}, 58(6):063001, September 1998.

\bibitem{NeterEtAl}
J.~Neter, M.~H. Kutner, C.~J. Nachtsheim, and W.~Wasserman.
\newblock {\em Applied linear statistical models}.
\newblock McGraw-Hill, New York, 4th edition, 1996.

\bibitem{RoeverMessengerPrix2011}
C.~R\"{o}ver, C.~Messenger, and R.~Prix.
\newblock Bayesian versus frequentist upper limits.
\newblock {\em Arxiv preprint 1103.2987}, February 2011.

\bibitem{AllenEtAl1999}
B.~Allen et~al.
\newblock Observational limit on gravitational waves from binary neutron stars
  in the galaxy.
\newblock {\em Physical Review Letters}, 83(8):1498--1501, August 1999.

\bibitem{AllenEtAl2005}
B.~Allen, W.~G. Anderson, P.~G. Brady, D.~A. Brown, and J.~D.~E. Creighton.
\newblock {Findchirp}: an algorithm for detection of gravitational waves from
  inspiraling compact binaries.
\newblock {\em Arxiv preprint gr-qc/0509116}, September 2005.

\bibitem{Brown2005}
D.~A. Brown.
\newblock Using the {I}nspiral program to search for gravitational waves from
  low-mass binary inspiral.
\newblock {\em Classical and Quantum Gravity}, 22(18):S1097--S1107, September
  2005.

\bibitem{WasEtAl2010}
M.~Was et~al.
\newblock On the background estimation by time slides in a network of
  gravitational wave detectors.
\newblock {\em Classical and Quantum Gravity}, 27(1):015005, January 2010.

\bibitem{Gosset1908}
W.~S. Gosset.
\newblock The probable error of a mean.
\newblock {\em Biometrika}, 6(1):1--25, March 1908.

\bibitem{KelejianPrucha1985}
H.~H. Kelejian and I.~R. Prucha.
\newblock Independent or uncorrelated disturbances in linear regression: {A}n
  illustration of the difference.
\newblock {\em Economics Letters}, 19(1):35--38, 1985.

\bibitem{BreuschRobertsonWelch1997}
T.~S. Breusch, J.~C. Robertson, and A.~H. Welsh.
\newblock The emperor's new clothes: a critique of the multivariate $t$
  regression model.
\newblock {\em Statistica Neerlandica}, 51(3):269--286, December 1997.

\bibitem{LangeEtAl1989}
K.~L. Lange, R.~J.~A. Little, and J.~M.~G Taylor.
\newblock Robust statistical modeling using the $t$~distribution.
\newblock {\em Journal of the American Statistical Association},
  84(408):881--896, December 1989.

\bibitem{Geweke1993}
J.~Geweke.
\newblock Bayesian treatment of the independent {S}tudent-$t$ linear model.
\newblock {\em Journal of Applied Econometrics}, 8:S19--S40, December 1993.

\bibitem{Divgi1990}
D.~R. Divgi.
\newblock Robust estimation using {S}tudent's $t$ distribution.
\newblock CNA Research Memorandum CRM~90-217, Center for Naval Analyses,
  Alexandria, VA, USA, December 1990.

\bibitem{McDonaldNewey1988}
J.~B. McDonald and W.~K. Newey.
\newblock Partially adaptive estimation of regression models via the
  generalized $t$ distribution.
\newblock {\em Econometric Theory}, 4(3):428--457, December 1988.

\bibitem{Hampel}
F.~R. Hampel, E.~M. Ronchetti, P.~J. Rousseeuw, and W.~A. Stahel.
\newblock {\em Robust statistics: The approach based on influence functions}.
\newblock Wiley, New York, 1986.

\bibitem{Huber}
P.~J. Huber and E.~M. Ronchetti.
\newblock {\em Robust statistics}.
\newblock Wiley, 2nd edition, 2009.

\bibitem{Creighton1999}
J.~D. Creighton.
\newblock Data analysis strategies for the detection of gravitational waves in
  non-{G}aussian noise.
\newblock {\em Physical Review~{D}}, 60(2):021101, July 1999.

\bibitem{AllenEtAl2002}
B.~Allen, J.~D.~E. Creighton, \'{E}.~\'{E}. Flanagan, and J.~D. Romano.
\newblock Robust statistics for deterministic and stochastic gravitational
  waves in non-{G}aussian noise: Frequentist analyses.
\newblock {\em Physical Review~{D}}, 65(12):122002, June 2002.

\bibitem{Allen2005}
B.~Allen.
\newblock $\chi^2$~time-frequency discriminator for gravitational wave
  detection.
\newblock {\em Physical Review~{D}}, 71(6):062001, March 2005.

\bibitem{DempsterLairdRubin1977}
A.~P. Dempster, N.~M. Laird, and D.~B. Rubin.
\newblock Maximum likelihood from incomplete data via the {EM} algorithm.
\newblock {\em Journal of the Royal Statistical Society~{B}}, 39(1):1--38,
  1977.

\bibitem{Wilks1938}
S.~S. Wilks.
\newblock The large-sample distribution of the likelihood ratio for testing
  composite hypotheses.
\newblock {\em The Annals of Mathematical Statistics}, 9(1):60--62, March 1938.

\bibitem{bspec2011}
C.~R\"{o}ver.
\newblock bspec: Bayesian spectral inference, version~1.3, 2011.
\newblock {R}~package. {URL}: \url{http://cran.r-project.org/package=bspec}.

\bibitem{MakelainenEtAl1981}
T.~M\"{a}kel\"{a}inen, K.~Schmidt, and G.~P.~H. Styan.
\newblock On the existence and uniqueness of the maximum likelihood estimate of
  a vector-valued parameter in fixed-size samples.
\newblock {\em The Annals of Statistics}, 9(4):758--767, July 1981.

\bibitem{DamourIyerSathyaprakash2001}
T.~Damour, B.~R. Iyer, and B.~S. Sathyaprakash.
\newblock Comparison of search templates for gravitational waves from binary
  inspiral.
\newblock {\em Physical Review~{D}}, 63(4):044023, January 2001.

\bibitem{AbbottEtAl2009a}
B.~P. Abbott et~al.
\newblock {LIGO}: the {L}aser {I}nter\-ferometer {G}ravi\-ta\-tion\-al-wave
  {O}bservatory.
\newblock {\em Reports on Progress in Physics}, 72(7):076901, July 2009.

\bibitem{Welch1967}
P.~D. Welch.
\newblock The use of {F}ast {F}ourier {T}ransform for the estimation of power
  spectra: A method based on time averaging over short, modified periodograms.
\newblock {\em {IEEE} Transactions on Audio and Electroacoustics},
  AU-15(2):70--73, June 1967.

\bibitem{TanakaTagoshi2000}
T.~Tanaka and H.~Tagoshi.
\newblock Use of new coordinates for the template space in a hierarchical
  search for gravitational waves from inspiraling binaries.
\newblock {\em Physical Review~{D}}, 62(8):082001, October 2000.

\bibitem{WilkGnanadesikan1968}
M.~B. Wilk and R.~Gnanadesikan.
\newblock Probability plotting methods for the analysis of data.
\newblock {\em Biometrika}, 55(1):1--17, March 1968.

\bibitem{Fawcett2006}
T.~Fawcett.
\newblock An introduction to {ROC} analysis.
\newblock {\em Pattern Recognition Letters}, 27(8):861--874, June 2006.

\bibitem{Waldman2011}
S.~Waldman.
\newblock {R}ayleigh distributions for {H1}, {L1} for {S6a}.
\newblock LIGO-Virgo collaboration internal report, November 2009.

\bibitem{Roever2011-LigoTechReport}
C.~R\"{o}ver.
\newblock Degrees-of-freedom estimation in the {S}tudent\mbox{-}$t$ noise
  model.
\newblock Technical Report LIGO-T1100497, LIGO-Virgo collaboration, September
  2011.

\bibitem{SlutskyEtAl2010}
J.~Slutsky et~al.
\newblock Methods for reducing false alarms in searches for compact binary
  coalescences in {LIGO} data.
\newblock {\em Classical and Quantum Gravity}, 27(16):165023, August 2010.

\bibitem{VeitchVecchio2008a}
J.~Veitch and A.~Vecchio.
\newblock Bayesian approach to the follow-up of candidate gravitational wave
  signals.
\newblock {\em Physical Review~{D}}, 78(2):022001, July 2008.

\end{thebibliography}

\end{document}